\documentclass[aip,jcp,reprint]{revtex4-1}

\usepackage[utf8x]{inputenc}
\usepackage{color}
\usepackage{placeins}

\usepackage[english,american]{babel}

\usepackage{blindtext}

\usepackage{amsmath,amsthm,amssymb}
\usepackage{bm}

\usepackage{graphicx}
\usepackage[caption=false]{subfig}

\usepackage{hyperref}

\usepackage[vlined, ruled]{algorithm2e}

\usepackage{multirow}

\begin{document}


\title{Estimation and uncertainty of reversible Markov models}

\author{Benjamin Trendelkamp-Schroer}

\thanks{Equal contribution}

\author{Hao Wu}

\thanks{Equal contribution}

\author{Fabian Paul}

\author{Frank Noé}

\email{frank.noe@fu-berlin.de}

\affiliation{Institut für Mathematik und Informatik, FU Berlin, Arnimallee 6,
14195 Berlin}

\date{\today}
\begin{abstract}
Reversibility is a key concept in Markov models and Master-equation
models of molecular kinetics. The analysis and interpretation of the
transition matrix encoding the kinetic properties of the model relies
heavily on the reversibility property. The estimation of a reversible
transition matrix from simulation data is therefore crucial to the
successful application of the previously developed theory. In this
work we discuss methods for the maximum likelihood estimation of transition
matrices from finite simulation data and present a new algorithm for
the estimation if reversibility with respect to a given stationary
vector is desired. We also develop new methods for the Bayesian posterior
inference of reversible transition matrices with and without given
stationary vector taking into account the need for a suitable prior
distribution preserving the meta-stable features of the observed process
during posterior inference. All algorithms here are implemented in
the PyEMMA software - \href{http://pyemma.org}{http://pyemma.org}
- as of version 2.0.\end{abstract}
\maketitle


\section{Introduction}

Markov models, Markov state models (MSMs), or Master-equation models
are a powerful framework to reduce the great complexity of bio-molecular
dynamics to a simple kinetic description that represents the underlying
transitions between distinct conformations \cite{schuette1999,swope2004,singhal2004,schultheis2005,noe2007,pan2008,prinz2011}.
These models allow us to analyze the longest-living (metastable) sets
of structures \cite{DeuflhardWeber_LAA05_PCCA+}, the effective transition
rates between them \cite{SinghalPande_JCP123_204909,NoeEtAl_PMMHMM_JCP13},
the kinetic relaxation processes and their relationship to equilibrium
kinetics experiments \cite{prinz2011,NoeEtAl_PNAS11_Fingerprints,Zhuang_JPCB11_MSM-IR,KellerPrinzNoe_ChemPhysReview11,LindnerEtAl_JCP13_NeutronScatteringI},
and the thermodynamics and kinetics over multiple thermodynamic states
\cite{SriramanKevrekidisHummer_JPCB109_6479,WuMeyRostaNoe_JCP14_dTRAM,WuNoe_MMS14_TRAM1,RostaHummer_DHAM}.
A key advantage of MSMs is that they are estimated from conditional
transition statistics between states, and they thus do not require
the data to be in global equilibrium across all states. As a result,
they are an excellent tool to integrate the data of multiple simulation
trajectories that have been run independently and from different initial
states into a single informative model \cite{ElmerParkPande_JCP123_114903,noe2009,BuchFabritiis_PNAS11_Binding}.

A variety of complex molecular processes have been successfully described
using MSMs. Examples include the folding of proteins into their native
folded structure \cite{noe2009,lane2011,BowmanVoelzPande_JACS11_FiveHelixBundle-TripletQuenching},
the dynamics of natively unstructured proteins \cite{perez2013,StanleyEstebanDeFabritiis_NatCommun14_KID},
and the binding of a ligand to a target protein \cite{held2011,BuchFabritiis_PNAS11_Binding,HuangCaflisch_PlosCB11_SmallMoleculeUnbinding,SilvaHuang_PlosCB_LaoBinding,bowman2012,PlattnerNoe_NatComm15_TrypsinPlasticity}.

There are two key steps in the construction of a MSM. At first a suitable
discretization of the continuous conformation space has to be obtained.
In most cases no good a-priori discretization is known and the discretization
has to be found based on the simulation data. The appropriate choice
of discretization is a topic of ongoing research \cite{chodera2007,perez2013,SchwantesPande_JCTC13_TICA}.
The error incurred by the discretization and by the subsequent approximation
of the jump-process as a Markov process can be systematically controlled
and evaluated \cite{sarich2010,prinz2011,NoeNueske_MMS13_VariationalApproach,NueskeEtAl_JCTC14_Variational}.

In the second step one estimates the transition probabilities between
pairs of states based on the transition statistics. The most common
approach to estimating Markov models from data is by means of a Bayesian
framework. One first harvests the transition counts, $c_{ij}(\tau)$
from the data, i.e. how often trajectories were found in discrete
state $i$ at some time $t$ and in discrete state $j$ at some later
time $t+\tau$. The parameter $\tau$ is called lag time and is crucial
for the quality of the Markov model \cite{sarich2010,prinz2011}.
Next, one computes the transition matrix either by maximizing the
likelihood, i.e. the probability over all possible Markov model transition
(or rate) matrices that may have generated the observed transition
counts \cite{BucheteHummer_JPCB08,bowman2009,prinz2011}; or by sampling
Markov models from the posterior distribution \cite{Singhal_JCP07,noe2008,ChoderaNoe_JCP09_MSMstatisticsII,SchuetteEtAl_JCP11_Milestoning,trendelkampnoe2013}.
A maximum likelihood estimate gives a single-point estimate, i.e.
a single Markov model that is ``most representative'' given the
data. However, if some transition events are rare compared to the
total simulation length - and this is the typical case in molecular
dynamics simulation - this maximum likelihood model might be very
uncertain and thus far away from the model that the one would converge
to by increasing amount of simulation data. The Bayesian posterior
ensemble is a natural approach to quantify such statistical uncertainties
and thus to make meaningful comparisons between a Markov models obtained
from different sets of simulations, or to experimental data.

A key property of molecular dynamics at thermal equilibrium, and a
necessary consequence of the second law of thermodynamics, is microscopic
reversibility of the equations of motion. This property is ensured
by many simulation procedures \cite{vanGunsteren1988,tuckerman1992}
and carries over to detailed balance between discrete states, i.e.
formally leads to a (time-) reversible Markov model. A reversible
Markov model is not only physically desirable, it offers statistical
advantages as it has only about half as many independent parameters
compared to a nonreversible model, and it allows the equilibrium kinetics
to be analyzed in a straightforward and meaningful manner. Furthermore
imposing detailed balance with respect to a given stationary vector
can be used to aid the efficient estimation of rare-event processes
from MSMs \cite{trendelkampnoe2014}. 

Algorithms imposing detailed balance during likelihood maximization
have been discussed in \cite{buchete-hummer:2008:coarse-master-equations,bowman2009,prinz2011,WuMeyRostaNoe_JCP14_dTRAM}.
First methods for the sampling the posterior distribution of reversible
transition matrices have been suggested in \cite{noe2008} and later
in \cite{MetznerNoeSchuette_Sampling}. A method working with natural
priors for reversible chains was proposed in \cite{bacallado2009}.
The sampling of transition matrices reversible with respect to a fixed
stationary distribution was also presented in \cite{noe2008}, while
a Gibbs sampling algorithm with a significantly improved convergence
rate has been developed in \cite{trendelkampnoe2013}. Ref. \cite{besag2013}
discusses methods for goodness-of-fit tests for Markov chains. 

This article is deliberately broad and presents new concepts, insights
and algorithms for reversible Markov model estimation in general,
maximum likelihood estimators and Bayesian estimators that mutually
benefit from each other. For this reason we first give a survey of
principles and consequences of reversible Markov models. We then extend
the framework of maximum likelihood estimation of transition matrices
by giving a simplified maximum likelihood estimator (MLE) for reversible
transition matrices and a new estimator for reversible transition
matrices with a fixed equilibrium distribution. The main part of the
paper comprises new algorithms for the full Bayesian analysis of the
posterior ensemble of reversible Markov models. As yet, three fundamental
problems have not been satisfactorily solved: (i) How can one harvest
statistically uncorrelated transition counts from trajectories in
which subsequent transitions are correlated, so as to give rise to
meaningful uncertainty intervals? (ii) Which prior should be used
in a Bayesian analysis so as to get error intervals that envelop the
 true value even for Markov models with many states? (iii) How can
we design efficient sampling algorithms for the reversible posterior
ensemble, i.e. algorithms that allow to quickly compute reliable error
bars for Markov models with many states? In this paper we discuss
(i) give a rather complete treatment of problems (ii) and (iii). Efficient
sampling algorithms are derived for reversible Markov models and reversible
Markov models with fixed equilibrium distribution.

\section{Reversible Markov models}

In this section we will show that microscopic reversibility carries
over to the discretized situation and discuss the desirable properties
of a reversible Markov state model.

\subsection{From microscopic reversibility to discrete-state detailed balance}

Let $\mu(x)$ denote the equilibrium distribution on the microscopic
degrees of freedom $x\in\Omega$, e.g. all-atom coordinates of the
system of interest, and let $p_{\tau}(x,y)$ denote the conditional
transition density of the MD implementation. $p_{\tau}(x,y)$ is the
probability that the system is found in state $y$ at time $t+\tau$
given that it has been in state $x$ at time $t$. The MD implementation
fulfills microscopic reversibility if the following detailed balance
equation 
\begin{equation}
\mu(x)\,p_{\tau}(x,\,y)=\mu(y)\,p_{\tau}(y,\,x)\label{eq:detailed_balance_microscopic}
\end{equation}
holds for all pairs of states $x,y\in\Omega$. Hence the terms ``detailed
balance'' and ``reversible'' are equivalent in our context. Since
$\mu(x)\,p_{\tau}(x,\,y)$ is the unconditional probability to find
the transition $(x,\,t)\rightarrow(y,\,t+\tau)$, Eq (\ref{eq:detailed_balance_microscopic})
means that the system is on average time-reversible - the absolute
number of transitions from $x$ to $y$ is equal to the reverse. Microscopic
detailed balance is desirable to have in any MD implementation when
the aim is to perform simulations in thermodynamic equilibrium. If
the (\ref{eq:detailed_balance_microscopic}) would be violated, that
would imply the existence of cycles $x\rightarrow y\rightarrow z\rightarrow x$
along which there is a net transport. Since such cycles could be used
to generate work, their existence in a system that is driven by purely
thermal energy would be inconsistent with the second law of thermodynamics.

Dynamical models that are commonly employed in MD implementations
fulfill detailed balance. Brownian (overdamped Langevin) dynamics
fulfills detailed balance. Hamiltonian and non-overdamped Langevin
dynamics fulfill generalized detailed balance with respect to momentum
inversion in phase space, but when integrating over the distribution
of momenta they do fulfill ordinary detailed balance in position space
\cite{koltai2014}. In practice, some finite time-stepping integrators
do not obey exact detailed balance with respect to the Boltzmann distribution,
but we here consider that the MD implementation has been chosen such
that detailed balance is at least approximately fulfilled.

Now suppose that the state space $\Omega$ is partitioned into non-overlapping
subsets $S_{1},\,...,\,S_{n}$ that we shall call \emph{discrete states}
here. Each set has an equilibrium probability given by 
\begin{equation}
\pi_{i}=\int_{S_{i}}\mathrm{d}x\,\mu(x)\label{eq:stationary_vector}
\end{equation}
and the transition density gives rise to a discrete state transition
matrix $P(\tau)$ with entries 
\begin{equation}
p_{ij}(\tau)=\frac{\int_{S_{i}}\mathrm{d}x\int_{S_{j}}\mathrm{d}y\,\mu(x)p_{\tau}(x,y)}{\int_{S_{i}}\mathrm{d}x\,\mu(x)}.\label{eq:transition_matrix}
\end{equation}
Using \eqref{eq:stationary_vector} and \eqref{eq:transition_matrix}
and microscopic detailed balance, \eqref{eq:detailed_balance_microscopic},
it is straightforward to verify that 
\begin{equation}
\pi_{i}p_{ij}(\tau)=\pi_{j}p_{ji}(\tau).\label{eq:detailed_balance_discretized}
\end{equation}
Note that (\ref{eq:detailed_balance_discretized}) holds independently
of the choice of the lag time $\tau$. Moreover, (\ref{eq:detailed_balance_discretized})
implies that $\pi$ is the equilibrium probability vector of $P(\tau)$.
By defining the diagonal matrix $\Pi=\mathrm{diag}(\pi_{1},\,...,\,\pi_{n})$,
we can alternatively write (\ref{eq:detailed_balance_discretized})
as a matrix equation:
\begin{align}
\Pi P & =(\Pi P)^{\top}\label{eq:DB_matrix}\\
P & =\Pi^{-1}P^{\top}\Pi\label{eq:DB_matrix2}
\end{align}
As a result, if the microscopic dynamics are reversible, the Markov
model transition matrix must also be reversible. However, a direct
estimate of $P$ from a finite amount of simulation data cannot be
expected to fulfill (\ref{eq:detailed_balance_discretized}) exactly.
Thus, the principle validity of detailed balance motivates us to \emph{enforce}
(\ref{eq:detailed_balance_discretized}) in the process of estimating
$P$.

Enforcing detailed balance helps to reduce the statistical error of
an estimator for $P$ because it reduces the number of independent
variables roughly by approximately one half (see Table \eqref{tab:dof}).
However, there are other consequences of having (\ref{eq:detailed_balance_discretized}):
Given detailed balance, we can compute the molecular equilibrium kinetics
in a physically meaningful way and employ some useful analysis tools
that are not defined for nonreversible Markov models. Moreover, we
can employ more efficient and robust matrix algebra routines when
exploiting that $P$ is a reversible matrix.

\begin{table}
\begin{centering}
\begin{tabular}{l|l|r}
constraints & dof & $\approx$\tabularnewline
\hline 
none & $n(n-1)$ & $n^{2}$\tabularnewline
\hline 
reversible & $\frac{1}{2}n(n-1)+n-1$ & $\frac{1}{2}n^{2}$\tabularnewline
\hline 
reversible, fixed $\pi$ & $\frac{1}{2}n(n-1)$ & $\frac{1}{2}n^{2}$\tabularnewline
\end{tabular}
\par\end{centering}

\protect\caption{\label{tab:dof}Number of independent variables (degrees of freedom,
dof) and their approximated values for transition matrices depending
on the constraints.}
\end{table}

\subsection{Eigenvalues and eigenvectors}

Many methods to analyze the molecular kinetics based on a Markov model
transition matrix rely on the eigenvalue decomposition of $P$. Using
the diagonal eigenvalue matrix $\Lambda=\mathrm{diag}(\lambda_{1},\,...,\,\lambda_{n})$
we can formulate a right eigenvalue problem with right column eigenvectors
$R=(r_{1},\,...,\,r_{n})$, $r_{i}\in\mathbb{R}^{n}$ and left row
eigenvectors $L=(l_{1},\,...\,l_{n})^{\top}$:
\begin{align}
PR & =R\Lambda\label{eq:eig_decomp_right}\\
LP & =\Lambda L.\label{eq:eig_decomp_left}
\end{align}
From (\ref{eq:eig_decomp_right}), we can obtain a generalized eigenvalue
problem:
\[
\Pi PR=\Pi R\Lambda.
\]
$\Pi$ is symmetric positive definite and as a result of detailed
balance, $\Pi P$ is symmetric. Hence, all eigenvalues $\lambda_{1},\,...,\,\lambda_{n}$
are real, and the eigenvectors are orthogonal with respect to the
equilibrium distribution \cite{Parlett_98_SymmetricEigenproblem}:
\[
\langle r_{i},\,r_{j}\rangle_{\pi}\propto\delta_{ij},
\]
where we have used the weighted scalar product $\langle u,v\rangle_{\pi}=\sum_{i}\pi_{i}u_{i}v_{i}$.
We can make $R$ orthonormal by scaling an arbitrarily obtained eigenvector
$r_{i}$ by $\langle r_{i},\,r_{i}\rangle_{\pi}^{-1/2}$.

Inserting the detailed balance formulation (\ref{eq:DB_matrix2})
into the decomposition (\ref{eq:eig_decomp_right}) immediately gives:
\begin{align*}
P^{\top}\Pi R & =\Pi R\Lambda\\
(\Pi R)^{\top}P & =\Lambda(\Pi R)^{\top}
\end{align*}
which is a left eigenvalue problem with the choice
\begin{align}
L & =(\Pi R)^{T}\nonumber \\
l_{i}^{\top} & =\Pi r_{i}.\label{eq:left-right-ev-relation}
\end{align}
Thus, detailed balance establishes a 1-to-1 relation between the left
and the right eigenvectors. We can decompose the transition matrix
into its spectral components by just using one set of eigenvectors
and the equilibrium distribution, such as:
\begin{equation}
P=R\mbox{\ensuremath{\Lambda}}R^{\top}\Pi=\sum_{i=1}^{n}\lambda_{i}r_{i}r_{i}^{\top}\Pi.\label{eq:P-decomposition}
\end{equation}

\paragraph{Example 1: }

\begin{figure}
\centering \includegraphics[width=0.8\columnwidth]{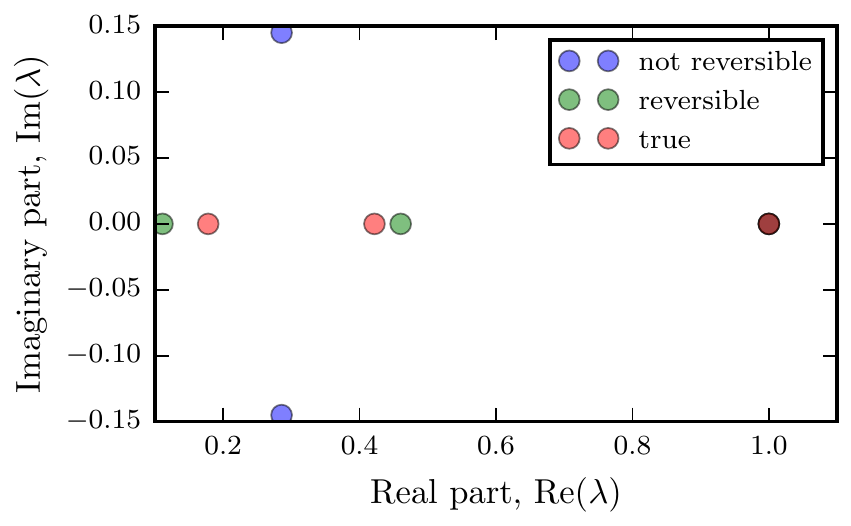}
\protect\caption{\label{fig:ev_3x3}Eigenvalues of $3\times3$ example system. The
eigenvalues obtained from the reversible estimate (green) are a closer
approximation to the true eigenvalues (red) than the eigenvalues obtained
from the non-reversible estimate (blue). The unique eigenvalue $\lambda=1$
is faithfully reproduced by both estimates.}
 
\end{figure}

Consider the following reversible $3\times3$ transition matrix,
\begin{equation}
P=\left(\begin{array}{ccc}
0.5 & 0.34 & 0.16\\
0.28 & 0.5 & 0.22\\
0.15 & 0.25 & 0.6
\end{array}\right).\label{eq:transition_matrix_3x3}
\end{equation}
Suppose we generate a Markov chain of length $20$ starting from state
1, resulting in the count matrix at lag $\tau=1$:
\begin{equation}
C=\left(\begin{array}{ccc}
4 & 3 & 0\\
1 & 4 & 3\\
1 & 1 & 2
\end{array}\right).\label{eq:count_matrix_3x3}
\end{equation}
Now we conduct a nonreversible and a reversible maximum likelihood
estimation of the transition matrix given $C$. Eigenvalues for the
exact transition matrix in \eqref{eq:transition_matrix_3x3} and both
nonreversible and reversible estimates for the given count matrix
in \ref{eq:count_matrix_3x3} are shown in Fig. \ref{fig:ev_3x3}.

It is seen that the nonreversible estimate contains complex eigenvalues.
These generally come in complex conjugate pairs. Fig. \ref{fig:ev_3x3}
shows a much higher accuracy of the reversible estimate compared to
the nonreversible estimate. In order to explore the statistical significance
of this observation, we run $N=1000$ chains of length $L=20$ using
transition matrix \eqref{eq:transition_matrix_3x3}. The reversible
and nonreversible estimation results, together with the true eigenvalues,
are reported below:

\noindent%
\begin{tabular}{c|c|l|l|l|l}
 & $\lambda_{1}$ & $\mathrm{Re}\{\lambda_{2}\}$ & $\mathrm{Im}\{\lambda_{2}\}$ & $\mathrm{Re}\{\lambda_{3}\}$ & $\mathrm{Im}\{\lambda_{3}\}$\tabularnewline
\hline 
exact & 1 & 0.42 & 0.0 & 0.18 & 0.0\tabularnewline
\hline 
rev & 1 & 0.36$\pm$0.31 & 0.0 & 0.18$\pm$0.04 & 0.0\tabularnewline
\hline 
nonrev & 1 & 0.32$\pm$0.29 & -0.04$\pm$0.08 & 0.07$\pm$0.21 & 0.04$\pm$0.09\tabularnewline
\end{tabular}

{\footnotesize{}}{\footnotesize \par}

It is seen that the reversible estimates do not only have the correct
real-valued structure, but can also have smaller uncertainties (here
especially for $\lambda_{3}$). This is expected to be a general result
due to the smaller number of degrees of freedom in the reversible
estimate.

\paragraph{Example 2: }

\begin{figure}
\centering \includegraphics[width=0.8\columnwidth]{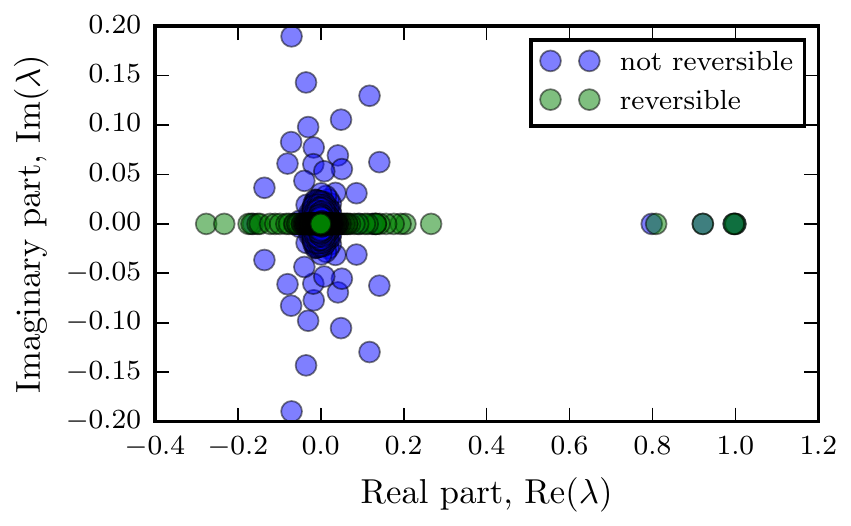}
\protect\caption{\label{fig:ev_alanine}Eigenvalues for alanine dipeptide. The cluster
of dominant eigenvalues indicates that the slowest processes are faithfully
reproduced by the non-reversible (blue) as well as by the reversible
(green) estimate. Only the eigenvalues of the reversible estimate
are purely real.}
 
\end{figure}

Fig. \ref{fig:ev_alanine} shows the distribution of eigenvalues from
nonreversible and reversible Markov models from simulation data for
the alanine-dipeptide molecule (see Sec. \ref{sub:applications}).
The eigenvalues of the reversible estimate are purely real while the
non-reversible estimate has eigenvalues with non-zero imaginary part.

\subsection{Equilibrium kinetics analyses}

Since detailed balance is a consequence of a system simulated at dynamical
equilibrium, it is not surprising that detailed balance in the transition
matrix $P$ is a prerequisite to analyze the equilibrium kinetics
given $P$. Since kinetics are related to slow processes we will here
only consider the $m$ largest eigenvalues $\lambda_{1},\,...\,\lambda_{m}$
and assume that they positive. Here are a few examples for equilibrium
kinetics properties computed from reversible transition matrices:
\begin{enumerate}
\item The dominant relaxation rates of the molecular system are:
\begin{equation}
\kappa_{i}=-\frac{1}{\tau}\ln\lambda_{i}\label{eq:relaxation-rate}
\end{equation}
where $i\ge2$ ($i=1$ has a relaxation rate of zero and corresponds
to the equilibrium distribution). The inverse quantities are the relaxation
timescales $t_{i}=\kappa_{i}^{-1}$. These rates or timescales are
of special interest because they are often detectable in kinetic experiments
such as fluorescence time-correlation spectroscopy, two-dimensional
IR spectroscopy or temperature jump experiments - see \cite{NoeEtAl_PNAS11_Fingerprints}
for a discussion.
\item The decomposition (\ref{eq:P-decomposition}) can be used to write
kinetic experimental observables in an illuminating form \cite{NoeEtAl_PNAS11_Fingerprints,KellerPrinzNoe_ChemPhysReview11,LindnerEtAl_JCP13_NeutronScatteringI}.
For example, the long-timescale part of the autocorrelation of a molecular
observable $a\in\mathbb{R}^{n}$, e.g. containing the fluorescence
values of every Markov state of a molecule, can be written as:
\begin{equation}
\mathrm{acf}(a;\,\tau)\approx\langle a,\,\pi\rangle^{2}+\sum_{i=2}^{m}\langle a,\,r_{i}\rangle_{\pi}^{2}\mathrm{e}^{-\kappa_{i}\tau}\label{eq:acf}
\end{equation}

\item The PCCA+ method for seeking $m$ metastable sets of Markov states
and its variants \cite{DeuflhardWeber_LAA05_PCCA+,RoeblitzWeber_AdvDataAnalClassif13_PCCA++}
assumes $m$ real-valued eigenvalues and eigenvectors. It is thus
only reliably applicable to reversible transition matrices.
\item Discrete transition path theory \cite{MetznerSchuetteVandenEijnden_TPT,NoeSchuetteReichWeikl_PNAS09_TPT,BerezhkovskiiHummerSzabo_JCP09_Flux}
computes the statistics of transition pathways from a set of states
$A$ to a set of states $B$ given a transition matrix. Discrete TPT
can be used with nonreversible and reversible transition matrices.
However, in the reversible case we get that the forward committor
and the backward committor are complementary:
\[
q_{i}^{+}=1-q_{i}^{-}
\]
and the net fluxes, when ordering states such that $q_{i}^{+}\le q_{j}^{+}$
are given by:
\[
f_{ij}^{+}=(q_{j}^{+}-q_{i}^{+})\pi_{i}p_{ij}
\]
which is analogous to an electric current $I=UG$ where $I=f_{ij}^{+}$,
$U=(q_{j}^{+}-q_{i}^{+})$ is the potential difference and $\pi_{i}p_{ij}$
is the conductivity \cite{EVandenEijnden_TPT_JStatPhys06}.
\end{enumerate}

\section{Likelihood, counting, and maximum likelihood estimation}

We restate the transition matrix likelihood and formulate the maximum
likelihood estimation problem for Markov model transition matrices.
We present new estimation algorithms for reversible Markov models
with known or unknown equilibrium distribution $\pi$.

\subsection{Likelihood }

Suppose we have a discrete sequence $S=\{s_{1},\,...,\,s_{N}\}$ with
$s_{i}\in\{1,\,...,\,n\}$. If we assume that this sequence is the
realization of a Markov chain with lag time $\tau=1$, the probability
that a transition matrix $P$ has generated $X$ is proportional to
the product of individual transition probabilities along the trajectory:

\begin{equation}
\mathbb{P}(S\mid P)\propto\prod_{t=1}^{N}p_{s_{t}s_{t+1}}\label{eq:likelihood-traj}
\end{equation}
We have neglected the proportionality constant because we won't need
them in order to maximize or sample from (\ref{eq:likelihood-traj}).
This is very handy because one component in this constant is the probability
of generating the first state, $p_{s_{0}}$, which is often unknown,
but is constant for a fixed data set $S$.

Suppose we have $c_{ij}$ transitions from $i$ to $j$. Then we can
group all $p_{ij}$ terms together and get a factor $p_{ij}^{c_{ij}}$.
Doing this for all pairs results in the equivalent likelihood formulation:
\begin{equation}
\mathbb{P}(C|P)\propto\prod_{i}\prod_{j}p_{ij}^{c_{ij}}\label{eq:likelihood-C}
\end{equation}
We can see that the count matrix $C\in\mathbb{R}^{n\times n}$ is
a sufficient statistics for the Markov model likelihood $\mathbb{P}(S\mid P)$
- it generates the same likelihood although we have discarded the
information in which sequence the transitions have occurred.

If multiple trajectories are available, their count matrices are simply
added up.

\subsection{Counting}

\label{sub:counting}

How should we count transitions for longer lag times $\tau>1$, or
if $S$ is not Markovian at lag time $\tau$? Regarding the first
case, if $S$ is Markovian at lag time $\tau$, a safe approach seems
to subsample the trajectory at time steps of $\tau$ and then treat
the subsampled trajectory as above \cite{SinghalPande_JCP123_204909,prinz2011}.
However this approach is statistically inefficient: If $S$ is also
Markovian for shorter lag times than $\tau$, then we are using less
information than we could. Even if $S$ only becomes Markovian at
lag times of $\tau$ or longer, transitions such as $1\rightarrow\tau+1$
and $\tau/2\rightarrow\tau+\tau/2$ are usually only partially correlated,
such that discarding the second transition is also not fully exploiting
the data. In practical MD simulations, the lag times required such
that a Markov model is a good approximation need to be quite long
(often in the the range of nanoseconds), such that subsampling the
data at $\tau$ will create severe problems with data and connectivity
loss. Regarding the second case, if $S$ is not Markovian at lag time
$\tau$ then treating every $c_{ij}$ as an independent count is incorrect.

Both cases can in principle be treated with the following formalism:
We always obtain the count matrix $c_{ij}$ in a \emph{sliding window}
mode \cite{prinz2011}, i.e. we harvest all $N-\tau$ available transition
counts from time pairs $(1\rightarrow\tau),\,(2\rightarrow\tau+1),\,...,\,(N-\tau\rightarrow N)$.
Unless $S$ is Markovian at lag time 1, we will now harvest more transition
counts than are statistically independent. We can formally correct
for this by introducing a statistical inefficiency $I_{ij}(\tau)$
for every count at a given lag time, such that $c_{ij}^{\mathrm{eff}}(\tau)=I_{ij}(\tau)\,c_{ij}(\tau)$
is the effective number of counts, resulting in the likelihood
\begin{equation}
\mathbb{P}(C|P)\propto\prod_{i,\,j}p_{ij}^{c_{ij}^{\mathrm{eff}}}.\label{eq:likelihood-SI-all}
\end{equation}
The determination of statistical inefficiencies for univariate signals
is well established \cite{Janke_NIC02_StatisticalAnalysis}. Determining
$I_{ij}(\tau)$ for transition count matrices is an open problem.
A first approach that allows for the first time to estimate consistent,
although somewhat too small uncertainty intervals for practical MD
data is discussed in \cite{Noe_preprint15_EffectiveCountMatrix}.
Note that the validity of the estimation algorithms described in the
present paper are independent of the choice of the count matrix, such
that future methods for estimating the effective count matrix can
be adopted without changing the estimation ablgorithms.

\subsection{Maximum likelihood estimation}

We will now assume that the effective counts are given. For better
readability we will subsequently omit the superscript $\mathrm{eff}$
and just use $C=(c_{ij})$ to indicate counts. Now we ask the question
what is the most likely transition matrix for the observation $C$,
i.e. we seek the maximum likelihood estimate (MLE) that maximizes
\eqref{eq:likelihood-traj} over the set of transition matrices.

\subsubsection{Non-reversible estimation}

It is well known that the non-reversible MLE for the transition probability
from state $i$ to state $j$ is simply given by the ratio of observed
counts from $i$ to $j$ divided by the total number of outgoing transitions
from state $i$ \cite{anderson1957}: 
\begin{equation}
\hat{p}_{ij}^{\mathrm{nonrev}}=\frac{c_{ij}}{\sum_{k}c_{ik}}.\label{eq:mle_nonreversible}
\end{equation}
We use the hat in order to denote an estimator. The term non-reversible
implies that reversibility has not been used as a constraint in the
estimation of $\hat{P}^{\mathrm{nonrev}}$. Of course $\hat{P}^{\mathrm{nonrev}}$
can be coincidentally reversible and will be reversible if the count
matrix $C$ is symmetric. For this reason, some early contributions
in the field forced symmetry in $C$ by counting $S$ forward and
backward. This practice is strongly discouraged as it will create
a large bias unless the trajectories used are very long compared to
the slowest timescales of the molecule.

\subsubsection{Reversible estimation}

Now we consider the problem of finding the reversible MLE $\hat{P}^{\mathrm{rev}}$
by enforcing detailed balance (\ref{eq:detailed_balance_discretized})
with respect to an unknown equilibrium distribution $(\pi_{i})$ as
constraint in the estimation procedure. Note that the count matrix
used for this approach is not modified, i.e. it comes from a forward-only
or nonreversible counting and is generally not symmetric. The constraints
(\ref{eq:detailed_balance_discretized}) can be more conveniently
handled by defining the new set of variables
\begin{equation}
x_{ij}=\pi_{i}p_{ij}.\label{eq:x_ij}
\end{equation}
Note that
\begin{equation}
x_{i}=\sum_{j}x_{ij}=\pi_{i}.\label{eq:X-rowsum}
\end{equation}
We can thus recover the transition matrix from $X=(x_{ij})$ by:
\begin{equation}
p_{ij}=\frac{x_{ij}}{x_{i}}.\label{eq:P-from-X}
\end{equation}
Inserting (\ref{eq:P-from-X}) into (\ref{eq:likelihood-C}) and adding
constraints for detailed balance and stochasticity leads to the reversible
maximum likelihood problem:
\begin{equation}
\begin{aligned}\underset{X}{\text{maximize}} &  & \sum_{i,j}c_{ij}\log\frac{x_{ij}}{\sum_{k}x_{ik}}\\
\text{subject to} &  & X_{ij}=X_{ji}\\
 &  & \sum_{k}X_{ik}>0\\
 &  & X_{ij}\geq0
\end{aligned}
\label{eq:mle_symmetric}
\end{equation}
Ignoring the inequality constraints the optimality conditions are
\begin{equation}
\frac{c_{ij}+c_{ji}}{x_{ij}}-\frac{c_{i}}{x_{i}}-\frac{c_{j}}{x_{j}}=0\label{eq:optimality_conditions}
\end{equation}
with $c_{i}=\sum_{j}c_{ij}$ and $x_{i}=\sum_{j}x_{ij}$. There is
no closed form solution when including the detailed balance constraint
so that \eqref{eq:mle_symmetric} has to be solved numerically. One
option is to directly solve \eqref{eq:optimality_conditions} for
$x_{ij}$ and turn it into a fixed-point iteration, as first proposed
in \cite{bowman2009}: 
\begin{equation}
x_{ij}^{(k+1)}=\frac{c_{ij}+c_{ji}}{\frac{c_{i}}{x_{i}^{(k)}}+\frac{c_{j}}{x_{j}^{(k)}}}\label{eq:update_symmetric}
\end{equation}
where $k$ counts the iteration number in the algorithm. For a starting
iterate $x_{ij}^{(0)}$ fulfilling the constraints in \eqref{eq:mle_symmetric},
for example $x_{ij}^{(0)}=(c_{ij}+c_{ji})/\sum_{i,j}(c_{ij}+c_{ji})$,
the iterates will be symmetric and fulfill the inequality constraints
for all $k>0$. 

If we sum over $j$ on both sides of \eqref{eq:update_symmetric}
and use \eqref{eq:X-rowsum}, we can instead reduce the problem to
iterative estimation of the equilibrium distribution:
\begin{equation}
\pi_{i}^{(k+1)}=\sum_{j=1}^{n}\frac{c_{ij}+c_{ji}}{\frac{c_{i}}{\pi_{i}^{(k)}}+\frac{c_{j}}{\pi_{j}^{(k)}}}\label{eq:fixed-point}
\end{equation}
The iteration is terminated when $\lvert\lvert\pi^{(k+1)}-\pi^{(k)}\vert\rvert<\epsilon$
. The final estimate $\hat{\pi}$ is then inserted into \eqref{eq:optimality_conditions}
to recover the reversible transition matrix estimate:
\begin{equation}
\hat{p}_{ij}^{\mathrm{rev}}=\frac{(c_{ij}+c_{ji})\pi_{j}}{c_{i}\pi_{j}+c_{j}\pi_{i}}\label{eq:P-given-pi}
\end{equation}
Note that both the optimum sought by Eq. (\ref{eq:fixed-point},\ref{eq:P-given-pi})
exhibits $\hat{p}_{ij}=0$ if $c_{ij}+c_{ji}=0$. Thus, in both optimization
algorithms, the sparsity structure of the matrix $\mathbf{C}+\mathbf{C}^{T}$
can be used in order to restrict all iterations to the elements that
will result in a nonzero element $\hat{p}_{ij}>0$.

\begin{figure}[H]
\begin{centering}
a) \includegraphics[width=0.8\columnwidth]{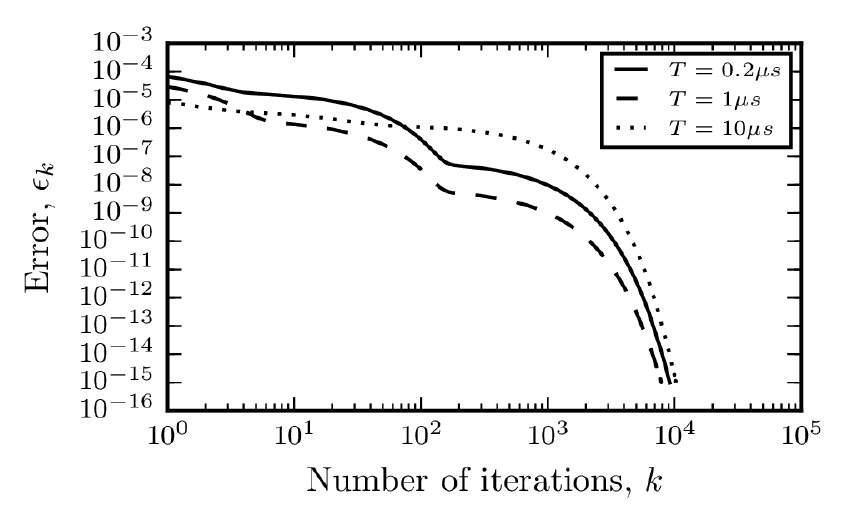}
\par\end{centering}

\begin{centering}
b) \includegraphics[width=0.8\columnwidth]{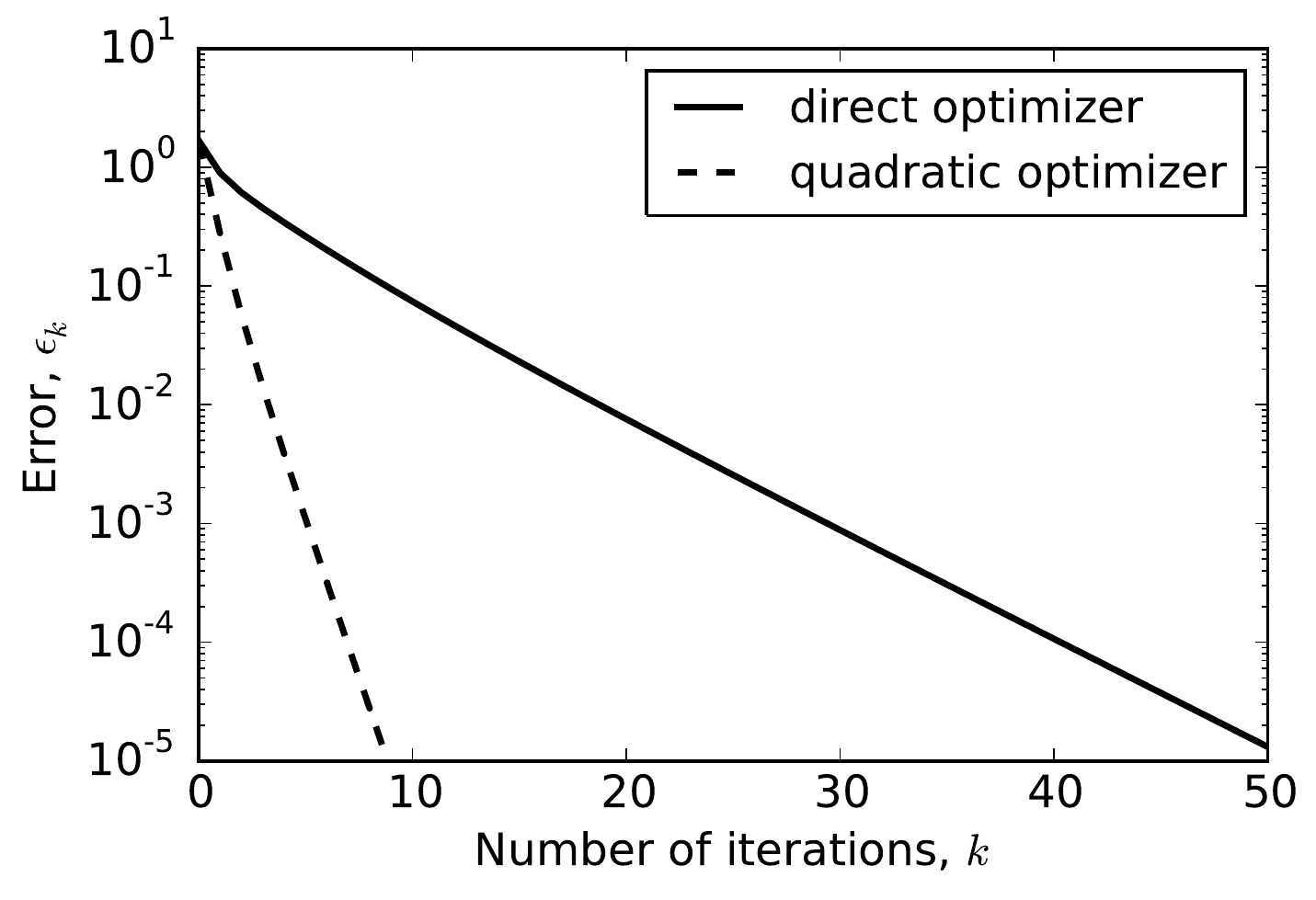}
\par\end{centering}

\centering{}c) \includegraphics[width=0.8\columnwidth]{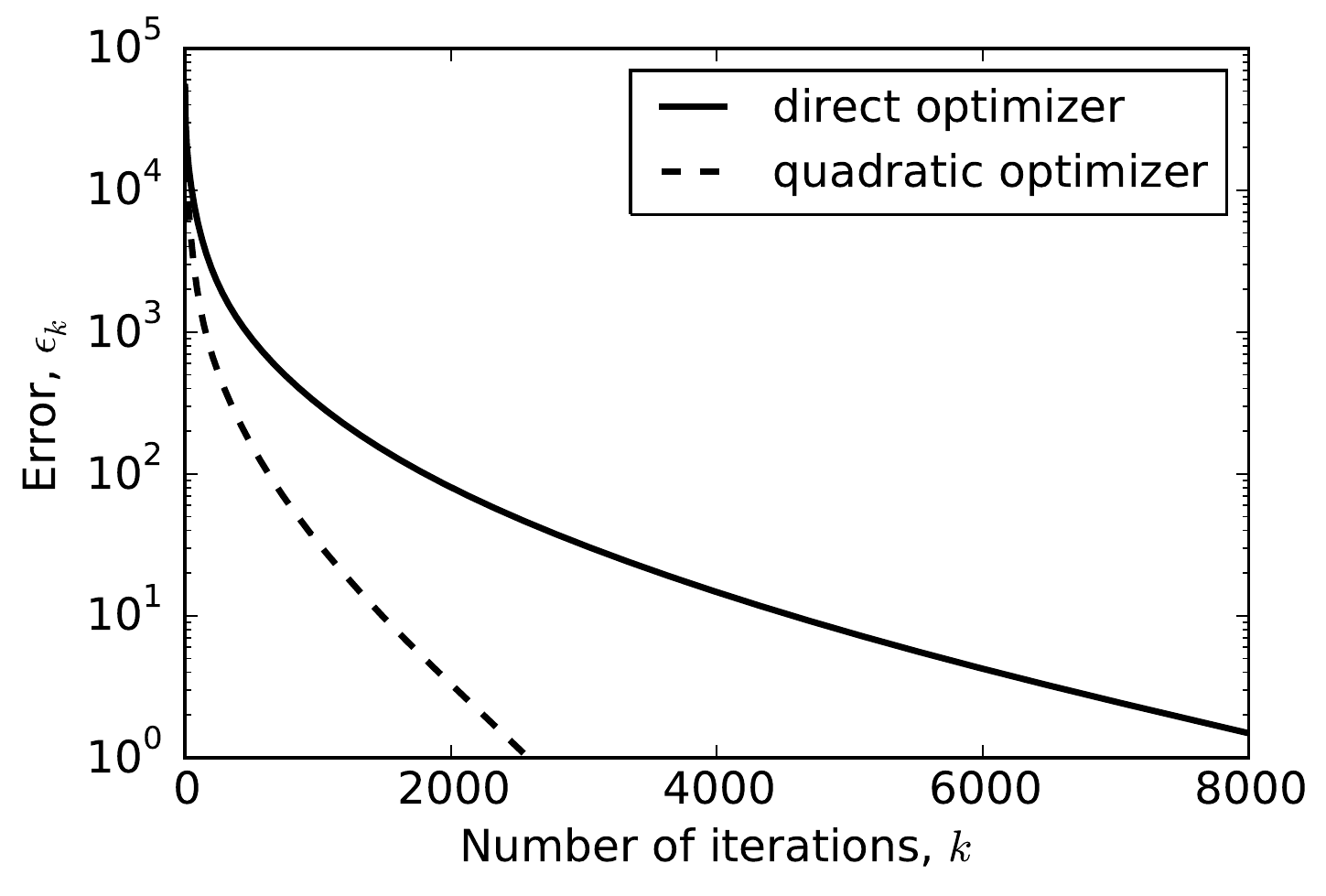}\protect\caption{\label{fig:mle_rev_iterative}Performance of algorithms for reversible
maximum likelihood estimation. (a) Reversible transition matrix estimated
using the fixed-point iteration \eqref{eq:fixed-point} from an $n=228$
state count-matrix obtained from alanine-dipeptide simulation data.
Convergence is shown for different total simulation lengths $T$.
(b) Performance comparison of the direct fixed-point iteration \eqref{eq:fixed-point}
and the quadratic optimizer described in Ref. \cite{prinz2011} for
reversible transition matrix estimation given the count matrix $C=\left((5,\,2,\,0),\,(1,\,1,\,1),\,(2,\,5,\,20)\right)^{\top}$.
Shown is the difference of the current likelihood to the optimal likelihood.
(c) Same as b, but using the 1734$\times$1734 count matrix from Pin
WW folding simulations used in Ref. \cite{NoeSchuetteReichWeikl_PNAS09_TPT}.}
 
\end{figure}

Furthermore, note that (\ref{eq:fixed-point},\ref{eq:P-given-pi})
are special cases of the transition-based reweighing analysis (TRAM)
method - see Ref. \cite{WuMeyRostaNoe_JCP14_dTRAM}, Eqs (29-30) -
for the special case of a single thermodynamic state. An example for
the progress of the self-consistent iteration using an alanine dipeptide
simulation is shown in Fig. \ref{fig:mle_rev_iterative}.

A different method of iterative solution presented in \cite{prinz2011}
updates $x_{ij}$ with the exact solution to the quadratic problem
arising from \eqref{eq:optimality_conditions} while holding all other
variables $x_{kl}$ fixed. As shown in Fig. \ref{fig:mle_rev_iterative},
this approach can exhibit faster convergence properties than the fixed-point
iteration \eqref{eq:fixed-point}.

\textbf{Uniqueness of the estimator}: The optimization problem \eqref{eq:mle_symmetric}
can be equivalently transformed into a convex optimization problem
by replacing the decision variables $x_{ij}$ with $z_{ij}=\log(x_{ij})$
(see \cite{WuNoe_MMS14_TRAM1} for details), which implies the uniqueness
of the maximum likelihood estimator.

\subsubsection{Reversible estimation for given stationary vector}

While unbiased MD simulations are useful to estimate state-to-state
transition probabilities $p_{ij}$, enhanced sampling algorithms such
as umbrella sampling and replica-exchange MD can be much more efficient
in order to gain insight of the equilibrium distribution $\pi$. Ref.
\cite{trendelkampnoe2014} demonstrates how an uncertain estimate
of $\pi$ can be combined with unbiased ``downhill'' trajectories
in order to estimate rare event kinetics. A key in such a procedure
is a way to estimate a reversible Markov model that is most likely
given transition counts observed from MD simulations, but at the same
time has a fixed equilibrium distribution $\pi$. Here we derive a
new, efficient estimation algorithm for this task. 

Enforcing reversibility with respect to a given stationary vector
results in the following constrained optimization problem 
\begin{align}
\underset{P}{\text{maximize}} &  & \sum_{i,j}c_{ij}\log p_{ij}\label{eq:mle_reversible_fixed_pi_likelihood}\\
\text{subject to} &  & \sum_{j}p_{ij}=1\label{eq:mle_reversible_fixed_pi_row}\\
 &  & p_{ij}\geq0\label{eq:mle_reversible_fixed_pi_ineq}\\
 &  & \pi_{i}p_{ij}=\pi_{j}p_{ji}.\label{eq:mle_reversible_fixed_pi_db}
\end{align}
$\pi$ can only be the unique stationary distribution of $P$ if $P$
is irreducible. To ensure irreducibility, we restrict the state space
to the largest (weakly) connected set of the undirected graph that
is defined by the adjacency matrix $C+C^{T}$. For a system with $n$
states, Eqs (\ref{eq:mle_reversible_fixed_pi_likelihood},\ref{eq:mle_reversible_fixed_pi_row},\ref{eq:mle_reversible_fixed_pi_ineq},\ref{eq:mle_reversible_fixed_pi_db})
 is a convex minimization problem in $\mathcal{\mathcal{{O}}}(n^{2})$
unknowns with $\mathcal{\mathcal{{O}}}(n^{2})$ equality and inequality
constraints. Solving this with a standard interior-point method requires
the solution of a linear system with $\mathcal{\mathcal{{O}}}(n^{2})$
unknowns to compute the search direction at each step. The resulting
computational effort of $\mathcal{\mathcal{{O}}}(n^{6})$ operations
for solving the linear system quickly becomes unfeasible for increasing
$n$. Therefore we will propose a fixed-point iteration that is also
feasible for large values of $n$.

\begin{figure}[H]
\centering \includegraphics[width=1\columnwidth]{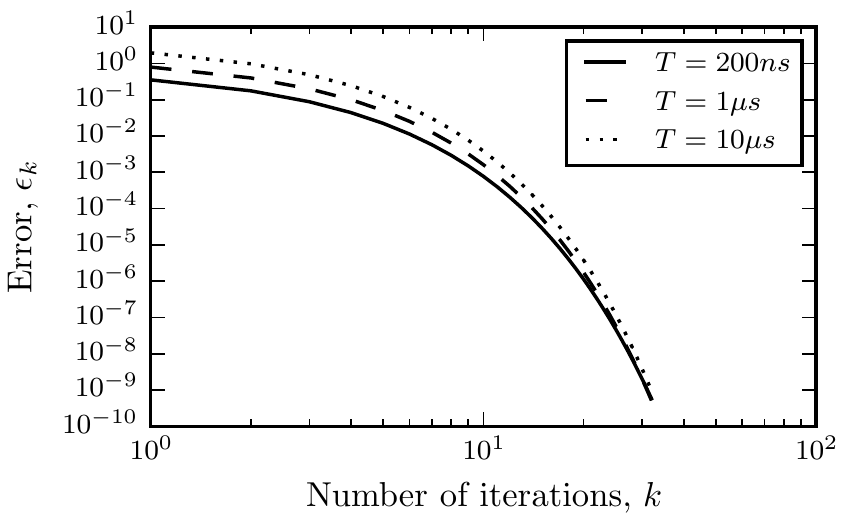}
\protect\caption{\label{fig:mle_rev_pi_iterative}Convergence of the reversible maximum
likelihood estimation with fixed stationary vector. The transition
matrix is estimated from an $n=228$ state count-matrix obtained from
alanine-dipeptide simulation data. Convergence is shown for different
total simulation lengths. The stationary distribution was obtained
using the simple counting estimate, \eqref{eq:stationary_vector_via_counting_frequencies}.}
 
\end{figure}

To solve the maximization problem, we ignore the inequality constraint
\eqref{eq:mle_reversible_fixed_pi_ineq} at first. The row-stochasticity
constraint \eqref{eq:mle_reversible_fixed_pi_row} is enforced by
introducing Lagrange multipliers $\lambda_{i}$ and adding penalty
terms $\lambda_{i}\left(\sum_{j}p_{ij}-1\right)$ for all $i=1,\,...,\,n$
to the objective function. The detailed balance constraint \eqref{eq:mle_reversible_fixed_pi_db}
is included into the likelihood explicitly by the change of variables
\[
p_{ij}^{\prime}=\begin{cases}
p_{ij} & \text{if }i\leq j\\
\frac{\pi_{j}}{\pi_{i}}p_{ji} & \text{else}
\end{cases}
\]
These substitutions result in the Lagrange function: 
\begin{align}
\begin{split}F=\sum_{i}c_{ii}\log p_{ii}^{\prime}+\sum_{i<j}(c_{ij}+c_{ji})\log p_{ij}^{\prime}\\
-\sum_{i<j}p_{ij}^{\prime}\left(\lambda_{i}+\lambda_{j}\frac{\pi_{i}}{\pi_{j}}\right)-\sum_{i}\lambda_{i}p_{ii}^{\prime}+\sum_{i}\lambda_{i}+const
\end{split}
\end{align}
that we seek to maximize. By setting the gradient of $F$ with respect
to all $p_{ij}^{\prime}$ to zero and subsequently reversing the change
of variables, we find the following expression for the maximum likelihood
estimate
\begin{equation}
\hat{p}_{ij}=\frac{(c_{ij}+c_{ji})\pi_{j}}{\lambda_{i}\pi_{j}+\lambda_{j}\pi_{i}}\label{eq:mle_reversible_fixed_pi_mle}
\end{equation}
Note the similarity of this equation with the maximum likelihood result
\eqref{eq:P-given-pi} where $\pi$ has been self-consistently computed
from the counts. The row counts $c_{i}$ are here replaced by the
yet unknown Lagrange multipliers $\lambda_{i}$. In order to find
the Lagrange multipliers, we sum Eq. (\ref{eq:mle_reversible_fixed_pi_mle})
over $j$: 
\begin{equation}
\sum_{j}\frac{(c_{ij}+c_{ji})\pi_{j}}{\lambda_{i}\pi_{j}+\lambda_{j}\pi_{i}}=1
\end{equation}
This doesn't give a closed-form expression for $\lambda_{i}$. However,
based on this equation, we propose the following fixed-point iteration
for the Lagrange multipliers:
\begin{equation}
\lambda_{i}^{(n+1)}=\sum_{j,\ c_{ij}+c_{ji}>0}\frac{(c_{ij}+c_{ji})\lambda_{i}^{(n)}\pi_{j}}{\lambda_{j}^{(n)}\pi_{i}+\lambda_{i}^{(n)}\pi_{j}}.\label{eq:mle_reversible_fixed_pi_iteration}
\end{equation}
Motivated by the analogy between Lagrange multipliers and row counts
described above, we set the starting point to:
\begin{equation}
\lambda_{i}^{(0)}=\frac{1}{2}\sum_{j}(c_{ij}+c_{ji})\label{eq:mle_reversible_fixed_pi_initial}
\end{equation}
Taking the limit $\lambda_{i}\rightarrow0^{+}$ in \eqref{eq:mle_reversible_fixed_pi_iteration}
still leads to a consistent solution. Choosing strictly positive starting
parameters according to \eqref{eq:mle_reversible_fixed_pi_initial}
results in valid iterates from \eqref{eq:mle_reversible_fixed_pi_iteration}.
In analogy to the reversible case we iterate \eqref{eq:mle_reversible_fixed_pi_iteration},
\eqref{eq:mle_reversible_fixed_pi_initial} until $\left\Vert \lambda^{(k+1)}-\lambda^{(k)}\right\Vert <\epsilon$.
An example for the progress of the self-consistent iteration using
alanine dipeptide simulation data is shown in Fig. \ref{fig:mle_rev_pi_iterative}.
Note that the converges is nearly three orders of magnitude faster
compared to the estimation with unknown equilibrium distribution (Fig.
\ref{fig:mle_rev_iterative}).

Given converged Lagrange multipliers, we can exploit \eqref{eq:mle_reversible_fixed_pi_mle}
to find the maximum likelihood transition matrix $\hat{P}$. For this
algorithm the inequality constraints \eqref{eq:mle_reversible_fixed_pi_ineq}
are automatically fulfilled when $c_{ij}\geq0$ for all $i,j$. Care
must be taken in two situations: (i) $\lambda_{i}=\lambda_{j}=0$
- one can show that the simultaneous limit $\lambda_{i}\rightarrow0^{+}$
and $\lambda_{j}\rightarrow0^{+}$ can only occur for $c_{ij}+c_{ji}=0$,
but then we know that $\hat{p}_{ij}=0$. (ii) A

 diagonal element $c_{ii}$ is zero. Depending on the values of $\pi$,
the solution $\lambda_{i}$ may take the value of zero such that equation
\eqref{eq:mle_reversible_fixed_pi_mle} for $i=j$ becomes $\hat{p}_{ii}=c_{ii}/\lambda_{i}=0/0$
which is meaningless and is not the correct limit of $p_{ii}$ as
$c_{ii}$ goes to zero. However this can be fixed easily by using
$\hat{p}_{ii}=1-\sum_{j\neq i}\hat{p}_{ij}$. for the diagonal elements
of $P$. In summary, we use the following equation for computing $\hat{P}$
from converged Lagrangian multipliers:
\begin{equation}
\hat{p}_{ij}=\begin{cases}
\pi_{j}\frac{c_{ij}+c_{ji}}{\lambda_{i}\pi_{j}+\lambda_{j}\pi_{i}} & i\neq j,\:\lambda_{i}+\lambda_{j}\neq0\\
0 & i\neq j,\:\lambda_{i}+\lambda_{j}=0\\
1-\sum_{j\neq i}\hat{p}_{ij} & i=j
\end{cases}
\end{equation}

\textbf{Uniqueness of the estimator}: Since the above estimation algorithm
is iterative, it is fair to ask whether the estimator $\hat{P}$ it
converges to is unique, or whether there might be multiple local maxima
that we could get stuck in. In this case, it is easy to show that
the estimator is unique: Let $P^{*}$ be an optimal transition matrix.
$p_{ij}^{*}=0$ exactly if $c_{ij}+c_{ji}=0$. Let $\Omega=\{p_{ij}|c_{ij}+c_{ji}>0\}$.
Then, the function $f(P)=\sum_{i,j}c_{ij}\log p_{ij}$ is strictly
convex on $\Omega$ and the constraints restrict the solution on a
convex subset $\tilde{\Omega}\subset\Omega$. The minimization of
a strictly convex function over a convex set has a unique solution.

\FloatBarrier

\section{Bayesian estimation \label{sec:posterior_ensemble}}

We introduce new algorithms for sampling the full posterior probability
distribution of Markov models, and in particular for estimating uncertainties
of quantities of interest, such as relaxation timescales or mean first
passage times. A key in these algorithms is the choice of a suitable
prior which enforces the sampled matrices to have the same sparsity
pattern as the transition count matrix, as this allows the credible
intervals to lie around the true value even for large transition matrices.
The relevance of the prior is first demonstrated for nonreversible
Markov models, for which an efficient sampling algorithm is known.
We then introduce new Gibbs sampling algorithms for reversible Markov
models with and without constraints on the equilibrium distribution
that vastly outperform previous algorithms for sampling reversible
Markov models.

\subsection{Bayes theorem and Monte Carlo sampling}

Bayes' formula relates the likelihood of an observed effective count
matrix $C$ given a probability model $P$ to the posterior probability
of the model given the observation, 
\begin{equation}
\underbrace{\mathbb{P}(P|C)}_{posterior}\propto\underbrace{\mathbb{P}(P)}_{prior}\underbrace{\mathbb{P}(C|P)}_{likelihood}.\label{eq:Bayes_formula}
\end{equation}
The posterior accounts for the uncertainty coming from a finite observation.
It incorporates a-priori knowledge about the quantity of interest
using the prior probability $\mathbb{P}(P)$. We will see that a suitable
choice of the prior is essential for the success of a Bayesian description
for high-dimensional systems.

In general, if we are interested in an observable that is a function
of a transition matrix, $f(P)$, we would like to compute its posterior
moments, such as the mean and the variance, 
\begin{align}
\langle f\rangle & =\int\mathrm{d}P\,\mathbb{P}(P|C)\,f(P),\label{eq:posterior_expectation}\\
\mathrm{Var}(f) & =\int\mathrm{d}P\,\mathbb{P}(P|C)\,\left(f(P)-\langle f\rangle\right)^{2}.\label{eq:posterior_variance}
\end{align}
While we usually use the maximum likelihood transition matrix $\hat{P}$
to provide ``best'' estimates, $f(\hat{P})$, the above integrals
are of interest because $\sigma(f)=\sqrt{\mathrm{Var}(f)}$ gives
us an estimate of the statistical uncertainty of $f$. Alternatively,
we might be interested in the credible intervals which encompass the
true value of $f$ with some probability, such as 0.683 ($1\sigma$
intervals) or 0.95 ($2\sigma$ intervals). As the integrals (\ref{eq:posterior_expectation},\ref{eq:posterior_variance})
are high-dimensional, we need to use Monte Carlo methods to approximate
them.

In Monte Carlo methods we generate a sample of transition matrices
$\{P^{(k)}\}_{k=1}^{N}$ distributed according to the posterior and
evaluate $f$ at each element $P^{(k)}$ in the ensemble. We then
approximate the posterior expectation value \eqref{eq:posterior_expectation}
and the posterior variance \eqref{eq:posterior_variance} by 
\begin{align}
m[f] & =\frac{1}{N}\sum_{k=1}^{N}f(P^{(k)}).\label{eq:approximation_posterior_expectation}\\
s^{2}[f] & =\frac{1}{N-1}\sum_{k=1}^{N}\left(f(P^{(k)})-m[f]\right)^{2}\label{eq:approximation_posterior_variance}
\end{align}
Obtaining good and reliable samples of the posterior $\mathbb{P}(P|C)$
is very difficult. Previous approaches have suffered from some or
all of the following difficulties, that are addressed here:
\begin{enumerate}
\item \textbf{Choice of the prior}: Given $n$ Markov states (typically
100s to 1000s), transition matrices have on the order of $n^{2}$
elements and are thus extremely high dimensional. Most priors used
in the past allow to populate all these elements $p_{ij}$, including
those for which no transition has been observed. Although the effect
of the prior can be overcome by enough simulation data, for any practical
amount of simulation data, such priors will lead to posterior distributions
whose probability mass is far away from the $\hat{P}$ true model.
This problem has been first addressed in Ref \cite{noe2009} by designing
a prior that equates mean and MLE for nonreversible transition matrices,
leading to credible intervals that nicely envelop the  the true value.
Here we design corresponding priors for reversible Markov models.
\item \textbf{Uncorrelated transition counts $C$}: As discussed in Sec.
\eqref{sub:counting}, the likelihood, and thus the posterior depends
on how transition counts are harvested from the discrete trajectories
which are generally time-correlated and not exactly Markovian at any
particular lag time $\tau$. While the MLE is often not or little
affected by the exact way of counting $C$, the uncertainties will
be dramatically different if $C$ is e.g. counted in a sliding window
mode (using transitions starting at all times $t=0,\,1,\,2,\,...$),
or by subsampling the trajectory (using transitions starting at all
times $t=0,\,\tau\,,2\tau$). Whereas the first approach underestimates
the uncertainties, the second approach often overestimates them and
is often not practical for large lagtimes $\tau$. Here we suggest
to use the effective number of uncorrelated transition counts, $C=C^{\mathrm{eff}}$,
and a first approach to compute them is described in Ref. \cite{Noe_preprint15_EffectiveCountMatrix}.
\item \textbf{Efficiency of the sampler}: Finally, given a choice of prior
and $C$, a sampling algorithm needs to explore the high-dimensional
space of transition matrices in a reasonable time. This is especially
problematic for reversible Markov models. The first Monte Carlo algorithm
for sampling the reversible posterior, described in Ref. \cite{noe2008},
suffers from poor mixing due to small acceptance probabilities of
the individual steps. In Ref. \cite{trendelkampnoe2013}, an improved
sampler was proposed. Here, we propose sampling algorithms for reversible
Markov models with and without fixed equilibrium distribution whose
efficiencies go far beyond previous approaches.
\end{enumerate}

\subsection{Non-reversible sampling}

Let us first illustrate the effect of prior choice on Bayesian estimation
of nonreversible Markov models. A convenient functional form for the
prior is the Dirichlet prior 
\begin{equation}
\mathbb{P}(P)\propto\prod_{i}\prod_{j}p_{ij}^{b_{ij}}\label{eq:Dirichlet_prior}
\end{equation}
where $B=(b_{ij})$ is a matrix of prior-counts. For this choice,
the posterior is given by 
\begin{equation}
\mathbb{P}(P|C)\propto\prod_{i}\left(\prod_{j}p_{ij}^{z_{ij}}\right).\label{eq:Dirichlet_posterior}
\end{equation}
$z_{ij}=c_{ij}+b_{ij}$ is the matrix of posterior pseudo-counts.
In the non-reversible case we can generate independent samples from
\eqref{eq:Dirichlet_posterior} by drawing rows of $P^{(k)}$ independently
from Dirichlet distributions $\prod_{j}p_{ij}^{\alpha_{ij}-1}$ with
Dirichlet parameters $\alpha_{ij}=z_{ij}+1=c_{ij}+b_{ij}+1$ \cite{singhal-pande:jcp:2005:adaptive-sampling}.

 Choosing a uniform prior, $b_{ij}=0$, assigns equal prior probability
to all entries, $p_{ij}$, in the posterior ensemble. But this a-priori
assumption can lead to serious problems when estimating quantities
for meta-stable systems. 

Consider for example the following transition matrix for a birth-death
chain consisting of two meta-stable sets $A=\{1,\dotsc,m\}$, $B=\{m+2,\dotsc,n\}$,
separated by a kinetic bottleneck in form of a single transition state,
\begin{equation}
P=\left(\begin{array}{ccccccccc}
\frac{1}{2} & \frac{1}{2} & 0\\
\frac{1}{2} & 0 & \frac{1}{2}\\
 & \ddots & \ddots & \ddots\\
 &  & 1-10^{-b} & 0 & 10^{-b}\\
 &  &  & \frac{1}{2} & 0 & \frac{1}{2}\\
 &  &  &  & 10^{-b} & 0 & 1-10^{-b}\\
 &  &  &  &  & \ddots & \ddots & \ddots\\
 &  &  &  &  &  & \frac{1}{2} & 0 & \frac{1}{2}\\
 &  &  &  &  &  &  & \frac{1}{2} & \frac{1}{2}
\end{array}\right).
\end{equation}
For barrier parameter $b=3$ and sets with $m=50$ and $n=101$ the
expected time for hitting $B$ from state $x=1$ is $2\cdot10^{5}$
steps. Now we are interested in the Bayesian estimator  for a simulation
of length $L=10^{7}$. The true distribution can be estimated with
arbitrary precision by repeating the simulation many times. Here,
$10^{3}$ repetitions led to an estimate of the 90\%  percentile for
the mean first passage time of $[1.5,\,2.7]\cdot10^{5}$ (see Table
below). 

In practice, we cannot afford to repeat the simulation many times
but would like to approximate the true value and its statistical uncertainty
from the given simulation data. Sampling the nonreversible posterior
given expected counts for a single chain of length $L=10^{6}$ with
a uniform prior, $b_{ij}=0$, results in non-zero transition probabilities
for elements $p_{ij}$ which are zero in the true transition matrix.
As a result artificial kinetic pathways circumventing the bottleneck
are appearing in the posterior ensemble which lead to a dramatic underestimate
of the mean first passage time.  The Bayesian estimate with 90\% credible
interval obtained from $10^{3}$ posterior samples is $[1.9,\,2.0]\cdot10^{3}$,
and thus two orders of magnitude smaller than the true value $2\cdot10^{5}$. 

Using the   prior $b_{ij}=-1$ suggested in Ref. \cite{noe2009} 
results in 90\% credible intervals, $[1.5,2.7]\cdot10^{5}$, which
clearly cover the true value $2\cdot10^{5}$. The choice $b_{ij}=-1$
leads to a posterior distribution in which sampled transition matrices
$P$ have the same sparsity structure as the count matrix $C$, i.e.
$p_{ij}=0$ if $c_{ij}=0$. As count matrices in the present context
are generally sparse, we call this prior briefly sparse prior. Apparently
the sparse prior leads to consistent credible intervals covering the
true value. 

\begin{center}
\begin{tabular}{l|r}
method & estimate \tabularnewline
\hline 
true &  $2.0_{1.5}^{2.7}\cdot10^{5}$\tabularnewline
\hline 
uniform prior $b_{ij}=0$ & $1.95_{1.9}^{2.0}\cdot10^{3}$\tabularnewline
\hline 
sparse prior $b_{ij}=-1$ & $2.0_{1.5}^{2.7}\cdot10^{5}$\tabularnewline
\end{tabular}
\par\end{center}

\begin{figure}
\includegraphics{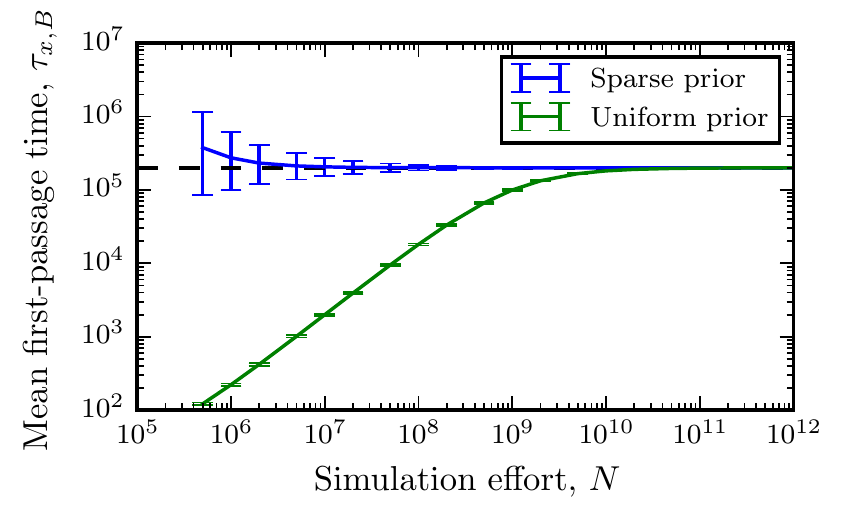}\protect\caption{Convergence of the 90\% credible interval for the sparse prior $b_{ij}=-1$
and the uniform prior $b_{ij}=0$. The dashed line indicates the true
value. The credible interval for the improper prior covers the true
value orders of magnitude before the credible interval for the uniform
prior. }

\label{fig:convergence_credint}
\end{figure}

Fig. \ref{fig:convergence_credint} shows the convergence of the 90\%
credible interval for the sparse and the uniform prior. The credible
interval for the sparse prior envelopes the true value already given
little data. To achieve consistency using the uniform prior requires
simulations order of magnitudes longer than the timescale of the slowest
process, thus rendering inference under this prior unpractical. 

Note that our prior induces a fixed sparsity structure. This concept
should not be confused with other sparsity inducing priors used i.e.
in the context of Bayesian compressed sensing \cite{ji2008bayesian},
where the sparsity pattern is subject to uncertainty.

\subsection{A prior for reversible Markov models}

Now we will present a new method for the sampling of reversible transition
matrices. In our new approach we replace the Dirichlet prior \eqref{eq:Dirichlet_prior}
by a new prior for reversible sampling.

Similarly as in reversible maximum likelihood estimation, we define
our reversible transition matrix sampler in the space of unconditional
transition probabilities $x_{ij}$. For convenience we restrict ourselves
to the independent set of variables with $i\le j$ (remember that
$x_{ij}=x_{ji}$ for reversible matrices), and keep them normalized
to 1:
\begin{align}
x_{ij} & \propto\pi_{i}p_{ij}\label{eq:transformation_P_X}\\
\sum_{i\geq j}x_{ij} & =1\label{eq:normalization_X}
\end{align}
Although $X$ is defined slightly different as in the maximum-likelihood
case, the mapping from $X$ back to $P$ is still given by Eq. (\ref{eq:P-from-X}).
We define a prior for reversible sampling on the set of $X$ matrices
rather than on $P$: Choosing $x_{ij}$ as the set of independent
variables has the advantage that obeying detailed balance amounts
to sampling symmetric matrices, $X=X^{T}$. 
\begin{equation}
\mathbb{P}(X)\propto\prod_{i\geq j}x_{ij}^{b_{ij}}.\label{eq:reversible_prior}
\end{equation}
The posterior for reversible sampling is then given by 
\begin{equation}
\mathbb{P}(X|C)\propto\prod_{i\geq j}x_{ij}^{b_{ij}}\prod_{i,j}\left(\frac{x_{ij}}{\sum_{k}x_{ik}}\right)^{c_{ij}}\label{eq:posterior_reversible}
\end{equation}
Below we will first consider how to sample from \eqref{eq:posterior_reversible}
using general prior counts $b_{ij}$. Then we will consider the specific
choice $b_{ij}=-1$ for all $i\le j$ and show that this choice has
similar properties as the sparse prior in the nonreversible case.

\subsection{Sampling reversible transition matrices\label{sub:Sampling-reversible-transition}}

There is no known method to generate independent samples from the
posterior under the reversibility requirement. Instead we will use
a a Markov chain Monte Carlo (MCMC) method to generate samples from
the posterior ensuring that each sampled transition matrix fulfills
the detailed balance condition \eqref{eq:detailed_balance_discretized}.
Our Markov chain will generate the ensemble $\{X^{(k)}\}_{k=1}^{N}$
via a set of updates advancing the chain from $X^{(k)}\rightarrow X^{(k+1)}$
starting from a valid initial state $X^{(0)}$. We can do a simple
row-normalization of the $X$ matrices to obtain the desired ensemble
$\{P^{(k)}\}_{k=1}^{N}$. Expectation values and variances will again
be estimated using (\ref{eq:approximation_posterior_expectation},\ref{eq:approximation_posterior_variance}).

Similarly as in \cite{noe2008,trendelkampnoe2013} we will construct
our Markov chain using a Gibbs sampling procedure, where we sample
a single element of $X$ in each step while leaving the other elements
unchanged. We repeat this sampling procedure for every element of
$X$, thus completing a Gibbs sweep. As detailed in the appendix,
we can use the following general Gibbs step to sample the posterior
\eqref{eq:posterior_reversible}:
\begin{enumerate}
\item Select an arbitrary element $x_{kl}$. Propose a new (unscaled) matrix
$X\rightarrow X'$ by sampling this element from the proposal density
$q(x_{kl}'|X)$:
\begin{equation}
x_{ij}'=\left\{ \begin{array}{cc}
\sim q(x_{kl}'|X) & (i,j)=(k,l)\\
x_{ij} & \text{else}
\end{array}\right.\label{eq:update_x_xprime-1}
\end{equation}
here $q(x_{kl}'|X)$ is an arbitrary, scale-invariant density. Scale-invariance
means that $q(x_{kl}'|X)\propto q(cx_{kl}'|cX)$ for any positive
constant $c$.
\item Accept $\bar{X}'$ as a new step in our Markov chain with probability
$\min\{1,p_{acc}\}$ where
\begin{align}
p_{acc} & =\left(1-x_{kl}+x_{kl}'\right)^{-\frac{n(n+1)}{2}-b_{0}}\frac{q(x_{kl}|X^{\prime})}{q(x_{kl}'|X)}\frac{\gamma(x_{kl}'|X)}{\gamma(x_{kl}|X')},\nonumber \\
\label{eq:p_acc-reversible}
\end{align}
where $b_{0}=\sum_{k\ge l}b_{kl}$ and $\gamma$ is the marginal density:
\[
\gamma(x_{kl}'|X)\propto\left\{ \begin{array}{ll}
\frac{(x_{kk}')^{c_{kk}+b_{kk}}}{(x_{k}-x_{kk}+x_{kk}')^{c_{k}}}, & k=l\\
\frac{(x_{kl}')^{c_{kl}+c_{lk}+b_{kl}}}{(x_{k}-x_{kl}+x_{kl}')^{c_{k}}(x_{l}-x_{kl}+x_{kl}')^{c_{l}}}, & k\neq l
\end{array}\right.
\]

\item Renormalize the matrix $X'\rightarrow\bar{X}'$ such that it fulfills
\eqref{eq:normalization_X}:
\begin{equation}
\bar{x}_{ij}^{\prime}=\frac{x_{ij}'}{1-x_{kl}+x_{kl}'}\label{eq:x-bar-prime}
\end{equation}

\end{enumerate}
While this approach will work for any choice of prior counts, we will
now use the sparse prior $b_{ij}=-1$ for all $i,j$ with the hope
to achieve similarly good results as in the nonreversible case. For
this choice, $\gamma(x_{kl}'|X)$ is scale-invariant, i.e. $\gamma(x_{kl}'|X)=\gamma(cx_{kl}'|cX)$,
and the Jacobian pre-factor in \eqref{eq:p_acc-reversible} is one.
Thus we have:
\begin{equation}
p_{acc}=\frac{\gamma(x_{kl}'|X)}{\gamma(x_{kl}|X')}\frac{q(x_{kl}|X^{\prime})}{q(x_{kl}'|X)}\label{eq:p_acc-rev-sparse}
\end{equation}
Thus the ideal choice of the proposal density is $q\equiv\gamma$,
which would guarantee that the acceptance probability is always $1$.
This proposal density degenerates to a point probability at zero if
$c_{kl}+c_{lk}=0$, which implies $b_{ij}=-1$ encodes a priori belief
that any transition for which neither the forward direction nor the
backward direction has ever been observed in the data has zero probability
in the posterior ensemble. Thus, this prior enforces $P$ to have
the same sparsity structure as the count matrix, like the choice $b_{ij}=-1$
for nonreversible sampling. Note that the reversible MLE has the same
sparsity structure as can be seen from the update rule \eqref{eq:update_symmetric}. 

We will choose proposal densities $\gamma(x_{kl}'|X)$ that are also
scale-invariant. In this case the normalization step 3 above ($X'\rightarrow\bar{X}'$)
can be omitted, i.e. if we accept $X'$, we can directly set it as
our new sample $X^{(k)}$ and obtain $P^{(k)}$ by row normalization.
We will now outline how to design the proposal density $\gamma$ such
that the acceptance probability is 1 or nearly 1. For $k=l$, sampling
$x_{kk}'\sim\gamma(x_{kk}'|X)$ is equivalent to sampling the following
transformed variable (see Appendix):
\begin{equation}
s^{\prime}=\frac{x_{kk}'}{x_{k}-x_{kk}+x_{kk}'}\sim\mathrm{Beta}(c_{kk},c_{k}-c_{kk})
\end{equation}
So we can simply define $q(x_{kk}'|X)\equiv\gamma(x_{kk}'|X)$ and
generate $x_{kk}'$ by
\begin{eqnarray}
s^{\prime} & \sim & \mathrm{Beta}(c_{kk},c_{k}-c_{kk})\nonumber \\
x_{kk}^{\prime} & = & (x_{k}-x_{kk})\frac{s^{\prime}}{1-s^{\prime}}\label{eq:y-sample-kk}
\end{eqnarray}
For $k\neq l$, there is no known way to draw independent samples,
but $\gamma(x_{kl}'|X)$ can be well approximated by a Gamma distribution
by matching the maximum point and the second derivative at the maximum.
A Gamma distribution can be efficiently sampled and we use it as proposal
density and accept the resulting $x_{kl}'$ with probability $\min\{1,p_{acc}\}$.
Specifically, our proposal step is:
\begin{equation}
x_{kl}'\sim\gamma(x_{kl}'|X)=\mathrm{Gamma}(x_{kl}'|\alpha,\,\beta)\label{eq:proposal_density}
\end{equation}
with the parameters
\begin{align}
\alpha & =-h\bar{v}\label{eq:alpha}\\
\beta & =-h\bar{v}^{2}\label{eq:beta}
\end{align}
using
\begin{align}
\bar{v} & =\frac{-b+\sqrt{b^{2}-4ac}}{2a}\label{eq:vbar}\\
h & =\frac{c_{k}}{(\bar{v}+x_{k}-x_{kl})^{2}}+\frac{c_{l}}{(\bar{v}+x_{l}-x_{kl})^{2}}-\frac{c_{kl}+c_{lk}}{\bar{v}^{2}}\label{eq:h}\\
a & =c_{k}+c_{l}-c_{kl}-c_{lk}\label{eq:a}\\
b & =(c_{k}-c_{kl}-c_{lk})(x_{l}-x_{kl})+(c_{l}-c_{kl}-c_{lk})(x_{k}-x_{kl})\label{eq:b}\\
c & =-(c_{kl}+c_{lk})(x_{k}-x_{kl})(x_{l}-x_{kl})\label{eq:c}
\end{align}
which matches the value and the first two derivatives of the true
marginal density at the maximum (see Appendix for derivation), and
leads to acceptance probabilities close to one for most values of
$x_{kl}$. However, if the current value of $x_{kl}$ is in one of
the heavy tails of the distribution $\gamma(x_{kl}'|X)$, the acceptance
probability can be much less than 1 and the Markov chain can get stuck.
In order to avoid this problem, we utilize a second step to generate
$x_{kl}'$: After we sample $x_{kl}'$ from the proposal density \eqref{eq:proposal_density}
we sample $x'_{kl}$ according to:
\begin{equation}
\log x_{kl}'\sim\mathcal{N}(\log x_{kl}'|\log x_{kl},1)
\end{equation}
where $\mathcal{N}(x|m,\,s)$ denotes the Normal distribution of $x$
with mean $m$ and standard deviation $s$. The proposal density defined
by the above update is
\begin{equation}
\tilde{q}(x_{kl}'|X)=\frac{1}{x_{kl}}\mathcal{N}(\log x_{kl}'-\log x_{kl}|0,1),
\end{equation}
and the corresponding acceptance probability is: 
\begin{equation}
p_{acc}=\frac{\gamma(x_{kl}'|X)}{\gamma(x_{kl}|X)}\frac{x_{kl}'}{x_{kl}}
\end{equation}

In summary, the proposed Algorithm \ref{alg:reversible_sampler} is
a Metropolis within Gibbs MCMC algorithm with adapted proposal probabilities
for each Gibbs sampling step. For efficiency reasons, transition matrix
elements $(i,\,j)$ for which no forward- or backward transition counts
have been observed, can be neglected in the sampling algorithm, in
order to account for the effect of the sparse prior.
\begin{algorithm}
\DontPrintSemicolon\SetAlgoNoLine\SetInd{0.5em}{1.0em}\KwIn{$C$,
$X^{(j)}$} \KwOut{$X^{(j+1)}$} \For {all indexes $(k,\,l)$
with $k\le l$ and $c_{kl}+c_{lk}>0$}{\If{$k=l$}{ Sample $x_{kk}^{(j+1)}$
from \eqref{eq:y-sample-kk}\; }\Else{ Calculate $\alpha$ and
$\beta$ by (\ref{eq:alpha},\ref{eq:beta}), and sample $x_{kl}'$
from $\mathrm{Gamma}(\alpha,\,\beta)$.\;$p_{acc}=\frac{\gamma(x_{kl}'|X)}{\gamma(x_{kl}|X)}\frac{\mathrm{Gamma}(x_{kl}|\alpha,\beta)}{\mathrm{Gamma}(x_{kl}'|\alpha,\beta)}$\;
Accept $x_{kl}'$ as $x_{kl}^{(j+1)}$ with probability $\min\{1,p_{acc}\}$\;
Sample $x_{kl}'$ by $\log x_{kl}'\sim\mathcal{N}(\log x_{kl},1)$.\;$p_{acc}=\frac{\gamma(x_{kl}'|X)}{\gamma(x_{kl}|X)}\frac{x_{kl}'}{x_{kl}}$\;
Accept $x_{kl}'$ as $x_{kl}^{(j+1)}$ with probability $\min\{1,p_{acc}\}$\;}}
\protect\caption{Reversible sampling algorithm \label{alg:reversible_sampler}}
\end{algorithm}

\subsection{A prior for reversible Markov models with fixed equilibrium distribution}

As before we will be working with variables $x_{ij}=\pi_{i}p_{ij}$
related to transition matrix entries $p_{ij}$ via \eqref{eq:transformation_P_X}.
In contrast to the previous algorithm, $\pi$ is not a function of
$P$ but fixed. Thus the single normalization condition \eqref{eq:normalization_X}
is replaced by a condition for each row: 
\begin{equation}
\sum_{j}x_{ij}=\pi_{i},\label{eq:normalization_fixed_pi}
\end{equation}
in order to ensure reversibility with respect to the given $\pi$.

All $x_{kl}$ in the lower triangle ($k>l$) are used as independent
variables. Given a valid $X$ matrix, an update that respects the
constraints is given by 
\begin{subequations}
\begin{align}
x_{kl} & \rightarrow x'_{kl}\label{eq:pi_conditional_sampling}\\
x_{kk} & \rightarrow x_{kk}+(x'_{kl}-x_{kl})\label{eq:pi_normalization1}\\
x_{lk} & \rightarrow x'_{kl}\label{eq:pi_symmetry}\\
x_{ll} & \rightarrow x_{ll}+(x'_{lk}-x_{lk}).\label{eq:pi_normalization2}
\end{align}
\eqref{eq:pi_normalization1} and \eqref{eq:pi_normalization2} ensure
that the normalization condition \eqref{eq:normalization_fixed_pi}
holds for the new sample and will thus keep $\pi$ constant, while
\eqref{eq:pi_symmetry} restores the symmetry and thus ensures reversibility
of $P$. We will again use the prior \eqref{eq:reversible_prior}
defined on the set of $X$ matrices and sample from the posterior
\eqref{eq:posterior_reversible}. The ideal proposal density of $x_{kl}^{\prime}$
is 
\begin{align}
\gamma(x_{kl}'\mid X)\propto & (x_{kl}')^{c_{kl}+c_{lk}+b_{kl}}\left(x_{kk}+x_{kl}-x_{kl}'\right)^{c_{kk}+b_{kk}}\nonumber \\
 & \left(x_{ll}+x_{kl}-x_{kl}'\right)^{c_{ll}+b_{ll}}\label{eq:conditional_fixed_pi}
\end{align}
which is the conditional distribution density for given all off-diagonal
elements of $X$ (except $x_{kl}$), $\pi$ and the counts $C$.
\end{subequations}

We have seen that a correct choice of prior parameters was essential
in order to successfully apply the posterior sampling for meta-stable
systems. As in the reversible case we will use $b_{kl}=-1$ for $k>l$
to enforce $x_{kl}=0$ whenever $c_{kl}+c_{lk}=0$. 

However, the choice of prior counts for the diagonal elements $b_{kk}$
is less straightforward. According to our experience, a good choice
is to determine the value of $b_{kk}$ based on the maximum likelihood
estimate $\hat{p}_{kk}$ of the $k$-th diagonal element as\textbf{
\begin{equation}
b_{kk}=\begin{cases}
0 & \:\:\:\:\:\hat{p}_{kk}>0,c_{kk}=0\\
-1 & \:\:\:\:\:\mathrm{else}
\end{cases}\label{eq:bkk-prior}
\end{equation}
}which ensures that the posterior expectation of $p_{kk}$ is zero
if and only if $\hat{p}_{kk}=0$, and the conditional expectation
of \eqref{eq:conditional_fixed_pi}, 
\begin{equation}
\mathbb{E}(x_{kl}'\mid X)=\frac{c_{kl}+c_{lk}}{c_{kl}+c_{lk}+c_{ll}}(x_{ll}+x_{kl})\label{eq:conditional_expectation_xkl-1}
\end{equation}
matches the MLE of the one-dimensional likelihood function for $x_{kl}$
given $X$ if $\hat{p}_{kk}>0$ and $c_{kk}=0$. (Note that for the
MLE of $P$, there is at most one $k$ which satisfies $\hat{p}_{kk}>0$
and $c_{kk}=0$ - see proof in Appendix.) 

However, in the case that $\hat{p}_{kk}=0$, the conditional \eqref{eq:conditional_fixed_pi}
would then degenerate so that $x_{kk}'=0$ with probability one, and
the $k$-th row and column of $X$ would remain fixed in the sampling
process. This effect can break ergodicity in the sampled Markov chain
and therefore prevent convergence of the algorithm. This problem is
avoided by regularizing the prior choosing the prior parameter as
$b_{kk}=-1+\epsilon$ for $\hat{p}_{kk}=0$ such that \eqref{eq:conditional_fixed_pi}
does not degenerate, where $\epsilon>0$ is a small number. In addition
we need to ensure that the Markov chain is started from an initial
state $X^{(0)}$ with $x_{kk}^{(0)}>0$. In summary, we select the
prior of $X$ for reversible sampling with fixed $\pi$ as\textbf{
\begin{equation}
b_{kk}=\begin{cases}
0 & \:\:\:\:\:\hat{p}_{kk}>0,c_{kk}=0\\
-1+\epsilon & \:\:\:\:\:\hat{p}_{kk}=0,c_{kk}=0\\
-1 & \:\:\:\:\:c_{kk}>0
\end{cases}\label{eq:bkk-prior-regularized}
\end{equation}
}

This choice of prior will again ensure that $c_{kl}+c_{lk}=0$ results
in $p_{kl}=0$ and $p_{lk}=0$ for all $k<l$ and for all posterior
samples, a property shared by the reversible MLE with fixed stationary
vector. This ensures that the posterior mass is located around the
maximum likelihood estimate $\hat{P}$ and again prevents the occurrence
of artificial kinetic pathways in the posterior ensemble.

\subsection{Sampling reversible Markov models with fixed equilibrium distribution}

We now investigate how to efficiently sample the conditional \eqref{eq:conditional_fixed_pi}.
Here we assume without loss of generality that $x_{kk}<x_{ll}$ and
transform $x_{kl}'\in(0,\,x_{kk}+x_{kl})$ into a new variable $v^{\prime}\in(0,+\infty)$
via 
\begin{equation}
v^{\prime}=\frac{x_{kl}'}{x_{kk}+x_{kl}-x_{kl}'}.\label{eq:transformation_x_v_fixed_pi}
\end{equation}
The ideal proposal density of $v^{\prime}$ is then
\begin{eqnarray}
\gamma_{v}(v^{\prime}|X) & \propto & \left|\frac{\partial x_{kl}'}{\partial v^{\prime}}\right|\gamma(x_{kl}'|X)\nonumber \\
 & = & \left(v^{\prime}\right)^{c_{kl}+c_{lk}+b_{kl}}\left(\frac{s}{s-1}+v^{\prime}\right)^{c_{ll}+b_{ll}}\nonumber \\
 &  & \cdot\left(1+v^{\prime}\right)^{-\left(c_{kl}+c_{lk}+c_{kk}+c_{ll}+b_{kl}+b_{kk}+b_{ll}+2\right)}\nonumber \\
 & = & \left(v^{\prime}\right)^{a_{1}}\left(\frac{s}{s-1}+v^{\prime}\right)^{a_{3}}\left(1+v^{\prime}\right)^{-\left(a_{1}+a_{2}+a_{3}\right)}
\end{eqnarray}
with
\begin{eqnarray*}
s & = & \frac{x_{ll}+x_{kl}}{x_{kk}+x_{kl}}\\
a_{1} & = & c_{kl}+c_{lk}+b_{kl}\\
a_{2} & = & c_{kk}+b_{kk}\\
a_{3} & = & c_{ll}+b_{ll}
\end{eqnarray*}
Like in the previous algorithm, we can approximate the conditional
of $v$ by a Gamma distribution as:

\begin{equation}
\gamma_{v}(v^{\prime}|X)\approx\mathrm{Gamma}(v^{\prime}|\alpha,\,\beta)\label{eq:proposal_density-1}
\end{equation}
with
\begin{align}
\alpha & =-h\bar{v}\label{eq:alpha-1}\\
\beta & =-h\bar{v}^{2}\label{eq:beta-1}
\end{align}
and
\begin{align}
\bar{v} & =\frac{-b+\sqrt{b^{2}-4ac}}{2a}\label{eq:vbar-1}\\
h & =-\frac{a_{1}}{\bar{v}^{2}}-\frac{a_{3}}{\left(\frac{s}{s-1}+\bar{v}\right)^{2}}+\frac{a_{2}}{\left(1+\bar{v}\right)^{2}}\label{eq:h-1}\\
a & =a_{2}+1\label{eq:a-1}\\
b & =a_{2}-a_{1}+\frac{a_{2}+a_{3}+1}{s-1}\label{eq:b-1}\\
c & =\frac{s\left(a_{1}+1\right)}{1-s}\label{eq:c-1}
\end{align}
See Appendix for derivation. Then $v^{\prime}$ can be sampled by
a Metropolis sampling step with the proposal density $\mathrm{Gamma}(v^{\prime}|\alpha,\beta)$
and the acceptance ratio
\[
\min\{1,p_{acc}\}=\min\left\{ 1,\frac{\gamma_{v}(v^{\prime}|X)\mathrm{Gamma}(v|\alpha,\beta)}{\gamma_{v}(v|X)\mathrm{Gamma}(v^{\prime}|\alpha,\beta)}\right\} 
\]
where $v=x_{kl}/x_{kk}$ denotes the original value of $v$. In addition,
in order to avoid the sampler from getting stuck at an extremely small
or large value of $v$, we also utilize the same strategy as in Section
\ref{sub:Sampling-reversible-transition} to generate $v^{\prime}$
by
\[
\log v^{\prime}\sim\mathcal{N}(\log v^{\prime}|\log v,1)
\]
. 

The proposed Algorithm \ref{alg:reversible_sampler_fixed_pi} for
sampling of reversible transition matrices with fixed stationary vector
can again be characterized as a Metropolis within Gibbs MCMC algorithm
with adapted proposal probabilities.

\begin{algorithm}
\DontPrintSemicolon \SetAlgoNoLine \SetInd{0.5em}{1.0em} \KwIn{$C$,
$\pi$, $X^{(j)}$} \KwOut{$X^{(j+1)}$}\For{all indexes $(k,\,l)$
with $k<l$ and $c_{kl}+c_{lk}>0$}{\If{$x_{kk}<x_{ll}$}{$v=\frac{x_{kl}}{x_{kk}}$
\;$s=\frac{x_{ll}+x_{kl}}{x_{kk}+x_{kl}}$\; $a_{1}=c_{kl}+c_{lk}+b_{kl}$\;$a_{2}=c_{kk}+b_{kk}$\;$a_{3}=c_{ll}+b_{ll}$\;}\Else{$v=\frac{x_{kl}}{x_{ll}}$\;$s=\frac{x_{kk}+x_{kl}}{x_{ll}+x_{kl}}$\;$a_{1}=c_{kl}+c_{lk}+b_{kl}$\;$a_{2}=c_{ll}+b_{ll}$\;$a_{3}=c_{kk}+b_{kk}$
\;}Calculate $\alpha$ and $\beta$ by \eqref{eq:alpha-1} and \eqref{eq:beta-1}\;
Sample $v^{\prime}$ from $\mathrm{Gamma}(\alpha,\beta)$.\;Let $x_{kl}^{\prime}=\min\{x_{kk}+x_{kl},x_{ll}+x_{kl}\}\cdot\frac{v}{1+v}$.\;$p_{acc}=\frac{\gamma_{v}(v^{\prime}|X)\mathrm{Gamma}(v|\alpha,\beta)}{\gamma_{v}(v|X)\mathrm{Gamma}(v^{\prime}|\alpha,\beta)}$
using \eqref{eq:pi_conditional_sampling}-\eqref{eq:pi_normalization2}\;Accept
$x_{kl}$ as $x_{kl}^{(j+1)}$ with probability $\min\{1,p_{acc}\}$\;Sample
$v'$ by by $\log v'\sim\mathcal{N}(\log v,1)$.\;$p_{acc}=\frac{\gamma_{V}(v'|X)v'}{\gamma_{V}(x_{kl}|X)v}$\;Let
$x_{kl}^{\prime}=\min\{x_{kk}+x_{kl},x_{ll}+x_{kl}\}\cdot\frac{v}{1+v}$.\;
Accept $x_{kl}'$ as $x_{kl}^{(j+1)}$ with probability $\min\{1,p_{acc}\}$\;}
\protect\caption{Reversible sampling algorithm with fixed stationary vector \label{alg:reversible_sampler_fixed_pi}}
\end{algorithm}

\clearpage

\section{Results}

\subsection{Validation}

\begin{figure}[H]
\begin{centering}
\includegraphics{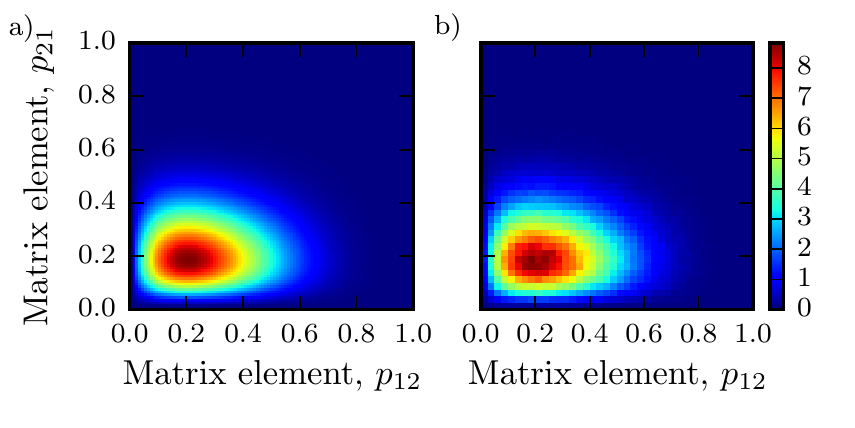}
\par\end{centering}

\protect\caption{\label{fig:sampled_vs_analytical_rev_2x2}Sampled histogram frequency
(a) and analytical probability density (b) of reversible posterior
for $2\times2$ count matrix. Sampled frequencies are in agreement
with the analytical probabilities.}
 
\end{figure}

We first demonstrate the validity of the reversible sampling algorithm
for the following $2\times2$ count-matrix, 
\begin{equation}
C=\left(\begin{array}{cc}
5 & 2\\
3 & 10
\end{array}\right).\label{eq:cmatrix_2x2}
\end{equation}
In Fig. \ref{fig:sampled_vs_analytical_rev_2x2} we compare the sampled
histograms using Algorithm \ref{alg:reversible_sampler} with analytical
values for the non-reversible posterior with Dirichlet-prior-counts
$b_{ij}=-1$. Any $2\times2$ transition matrix automatically fulfills
detailed balance, and therefore the analytical and sampled densities
are expected to be equal. The histogram for samples of the reversible
algorithm are indeed in agreement with the analytical posterior.

\begin{figure}[H]
\centering{}\includegraphics[width=1\columnwidth]{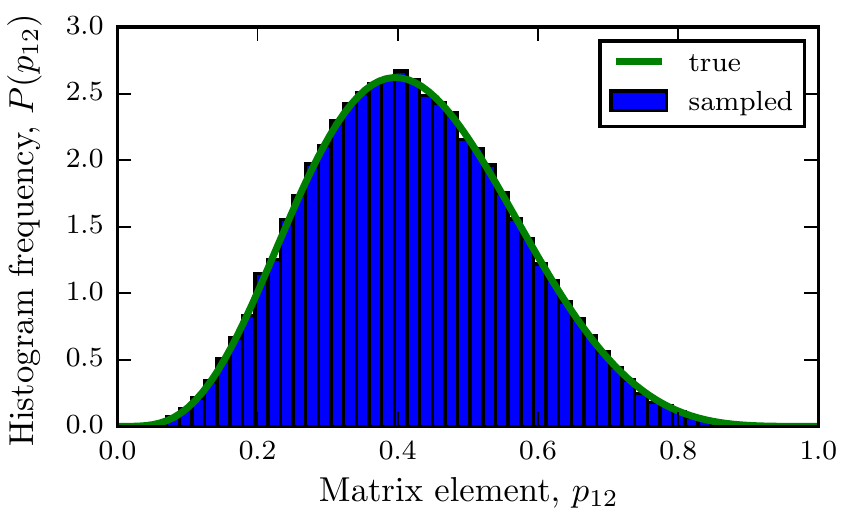}
\protect\caption{\label{fig:sampled_vs_analytical_revpi_2x2}Sampled histograms and
analytical probability density for reversible posterior with fixed
stationary vector for $2\times2$ count-matrix.}
 
\end{figure}

In Fig. \ref{fig:sampled_vs_analytical_revpi_2x2} the sampled histogram
for count matrix \eqref{eq:cmatrix_2x2} with fixed stationary distribution
$\pi=\left(0.25,\,0.75\right)^{\top}$ using Algorithm \ref{alg:reversible_sampler_fixed_pi}
is compared with the exact posterior distribution. The detailed balance
relation \eqref{eq:detailed_balance_discretized} with fixed stationary
vector enforces a linear dependency between the transition matrix
element $p_{12}$ and $p_{21}$. The resulting posterior is therefore
restricted to the line $\pi_{1}p_{12}=\pi_{2}p_{21}$ such that the
projection on $p_{12}$ in Fig. \ref{fig:sampled_vs_analytical_revpi_2x2}
already contains the full information about the one-dimensional posterior.
A comparison between histogram frequency and analytical density demonstrates
the validity of the algorithm. 

\FloatBarrier

\subsection{Applications}

\label{sub:applications}

To demonstrate the usefulness of the proposed algorithms we apply
them to molecular dynamics simulation data. Here, two systems are
chosen to illustrate our methods: (1) The alanine-dipeptide molecule,
and (2) the bovine pancreatic trypsin inhibitor molecule (BPTI). 

We start by discussing the alanine dipeptide results. The system was
simulated on GPU-hardware using the OpenMM simulation package \cite{eastman2013}
using the \emph{amber99sb-ildn} forcefield \cite{lindorff2010} and
the \emph{tip3p} water model \cite{jorgensen1983}. The cubic box
of length $2.7nm$ contained a total of $652$ solvent molecules.
We used Langevin equations at $T=300K$ with a time-step of $2fs$.
A total of $10\mu s$ of simulation data was generated. The $\phi$
and $\psi$ dihedral angles were discretized using a $20\times20$
regular grid to obtain a matrix of transition counts $C=(c_{ij})$,
here by sampling one count per lag time $\tau$. Below we will show
histograms for two important observables, \emph{largest implied time-scales}
$t_{i}$ and \emph{expected hitting times}, $\tau(A\rightarrow B)$,
for pairs $A$, $B$ of meta-stable sets. We compute the posterior
sample-mean and 90\% credible intervals for $1\mu s$ of simulation
data and  show that the credible intervals nicely envelop a reference
value obtained from the MLE transition matrix for the total simulation
data, supporting  the proposed prior as a 'good' choice for reversible
sampling in meta-stable systems.

\subsubsection{Alanine dipeptide, reversible sampling}

\begin{figure*}
(a)\includegraphics[width=0.25\textwidth]{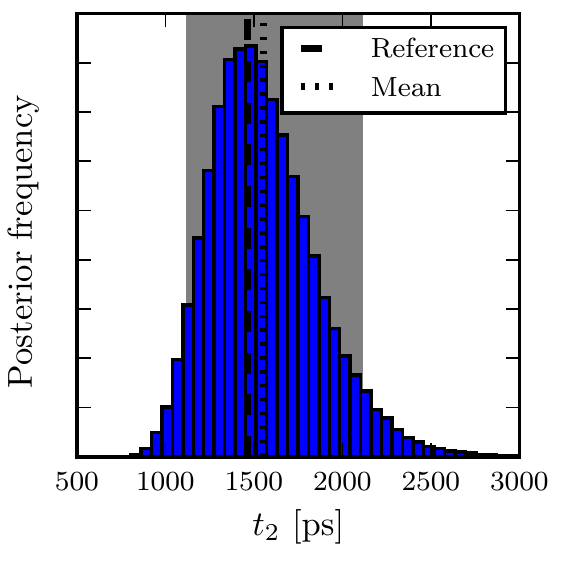} (b)\includegraphics[width=0.25\textwidth]{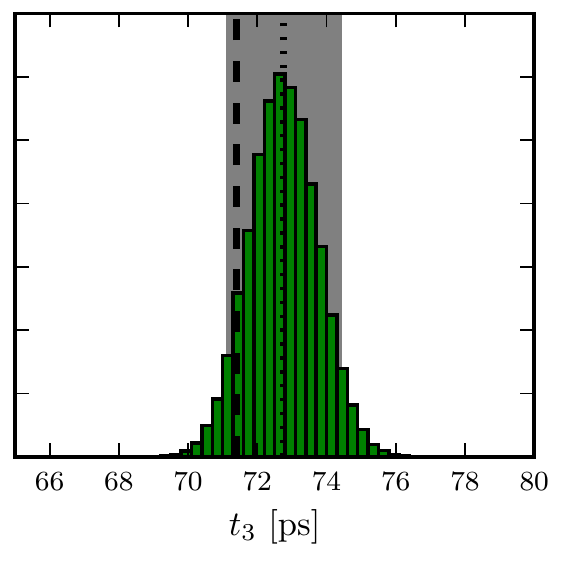}
(c)\includegraphics[width=0.25\textwidth]{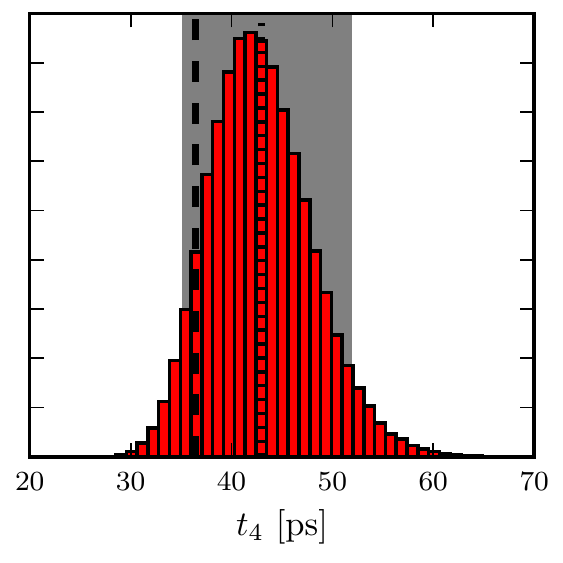}

(d)\includegraphics[width=0.25\textwidth]{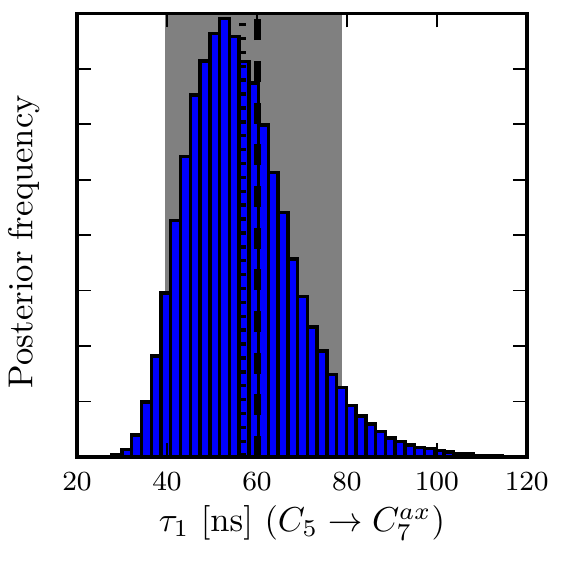} (e)\includegraphics[width=0.25\textwidth]{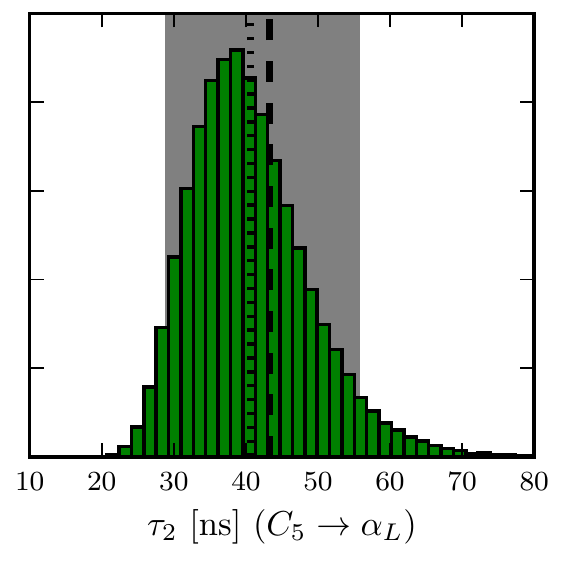}
(f)\includegraphics[width=0.25\textwidth]{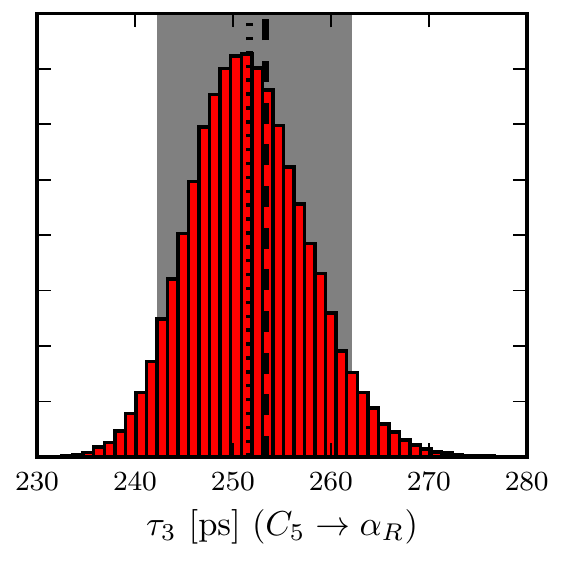}\protect\caption{\label{fig:tsrev}a)-c) Implied time-scales, $t_{i}$. Histograms
obtained from reversible posterior sampling. Dashed lines indicate
the  reference value, $\hat{t}_{i}$, dotted lines indicate the posterior
sample mean, $\mu(t_{i})$. MLE and posterior mean are in very good
agreement for the proposed choice of prior. The 90\% credible intervals
are the shaded regions in gray. Expected hitting times, $\tau$, d)-f).
Histograms obtained from reversible posterior sampling. Dashed lines
indicate the  reference value, $\hat{\tau}$, dotted lines indicate
the posterior sample mean, $\mu(\tau)$.  In all cases the reference
value obtained from a long simulation is clearly compatible with the
posterior sample (credible interval).}
\end{figure*}

In Fig. \ref{fig:tsrev} we show histograms of implied time-scales
computed from a reversible posterior sample. The mean values estimated
from the posterior sample are in good agreement with the reference
values. Tables \ref{tab:ts_rev} and \ref{tab:mfpt_rev} compare the
 reference values with the sample mean $\mu$ and sample standard
deviation $\sigma$ for each observable. 

In order to gain a first impression of the efficiency of the sampling
and the quality of our estimates, we compute the integrated autocorrelation
time $t_{corr}$ for each quantity sampled (here implied timescales
and hitting times). The error of the sample mean $m[f]$ compared
to the true mean $\langle f\rangle$ can then be estimated as 
\begin{equation}
\epsilon=\mathbb{E}[(m[f]-\langle f\rangle)^{2}]=\frac{\sigma^{2}[f]}{N_{\mathrm{eff}}}\label{eq:MCMC_error-1}
\end{equation}
where $N_{\mathrm{eff}}=N/(1+2t_{corr})$ is the effective number
of samples with $N$ the total number of samples. See Ref. \cite{geyer1992}
for a thorough discussion. $t_{corr}$ and $\epsilon$ are also reported
in Table \eqref{tab:ts_rev}.

\begin{table}
\begin{ruledtabular}
\begin{tabular}{rrrrrr}
 & $\hat{t}_{i}/ps$  & $\mu/ps$  & $\sigma/ps$  & $\epsilon/ps$  & $t_{corr}$ \tabularnewline
$t_{2}$  & 1462 & 1556  & 303  & 19.00  & 197 \tabularnewline
$t_{3}$  & 71  & 73  & 1  & 0.01  & 10 \tabularnewline
$t_{4}$  & 36  & 43  & 5  & 0.06  & 7 \tabularnewline
\end{tabular}
\end{ruledtabular}

\protect\caption{\label{tab:ts_rev}Comparison of reference implied time-scales ($\hat{t}_{i}$)
 with mean $\mu$ and standard deviation $\sigma$ from the reversible
posterior using $N=10^{5}$ samples. $\epsilon$ is the estimated
error of the mean $\mu$ and $t_{\mathrm{corr}}$ is the autocorrelation
time of the sampled quantity.}
 
\end{table}

Fig. \ref{fig:tsrev}d-f) show histograms for expected hitting times
for the three transitions $C_{5}\rightarrow C_{7}^{ax}$, $C_{5}\rightarrow\alpha_{L}$
and $C_{5}\rightarrow\alpha_{R}$ between meta-stable sets in the
$\phi$ and $\psi$ dihedral angle plane. Again, mean values are in
good agreement with the corresponding  reference values. Table \eqref{tab:mfpt_rev}
summarizes the computed results. The table columns again contain the
reference value $\hat{\tau}$ the mean value $\mu$, the standard
deviation $\sigma$, the estimated correlation time $t_{corr}$ and
the error of the mean value, $\epsilon$. 

\begin{table}
\begin{ruledtabular}
\begin{tabular}{rrrrrr}
 & $\hat{\tau}/ns$  & $\mu/ns$  & $\sigma/ns$  & $\epsilon/ns$  & $t_{corr}$ \tabularnewline
$\tau(C_{5}\rightarrow C_{7}^{ax})$  & 60.4 & 56.7  & 12.0  & 0.77  & 202 \tabularnewline
$\tau(C_{5}\rightarrow\alpha_{L})$  & 43.6  & 40.6  & 8.3  & 0.53  & 206 \tabularnewline
$\tau(C_{5}\rightarrow\alpha_{R})$  & 0.253  & 0.250  & 0.005  & 0.0004  & 218 \tabularnewline
\end{tabular}
\end{ruledtabular}

\protect\caption{\label{tab:mfpt_rev}Expected hitting times computed from reversible
posterior using $N=10^{5}$ samples. See Table \ref{tab:ts_rev} for
definition of other symbols.}
 
\end{table}

\subsubsection{Alanine dipeptide, reversible sampling with fixed equilibrium distribution}

Below we report results for sampling with fixed stationary distribution.
The stationary distribution $\pi_{i}$ was, for sake of simplicity,
computed using the relative frequencies of state occurrences, 
\begin{equation}
\pi_{i}=\frac{\sum_{k}c_{ik}}{\sum_{j,k}c_{jk}}.\label{eq:stationary_vector_via_counting_frequencies}
\end{equation}
It should be noted that a more useful and independent source of $\pi$
are enhanced sampling simulations targeted at rapidly generating a
good estimate of the equilibrium probabilities alone. See Ref. \cite{trendelkampnoe2014}
for methods and applications that combine MD simulations and enhanced
sampling simulations in order to efficiently compute rare-event kinetics. 

\begin{figure*}
(a)\includegraphics[width=0.25\textwidth]{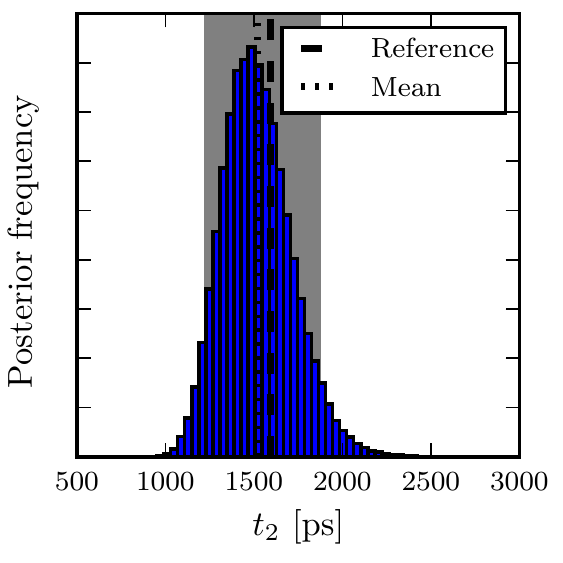} (b)\includegraphics[width=0.25\textwidth]{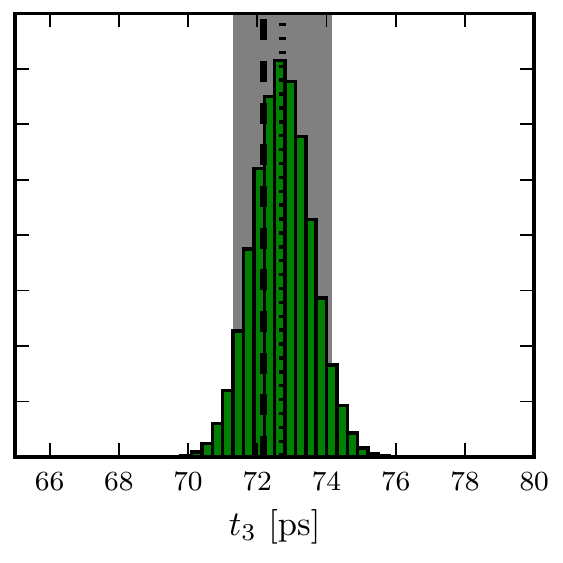}
(c) \includegraphics[width=0.25\textwidth]{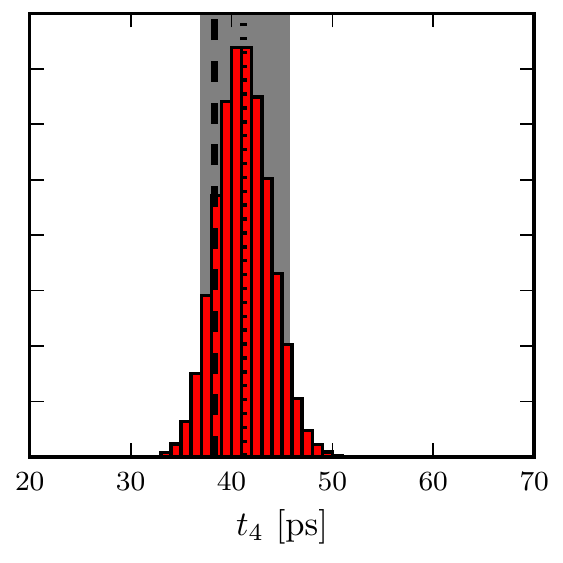}

(d) \includegraphics[width=0.25\textwidth]{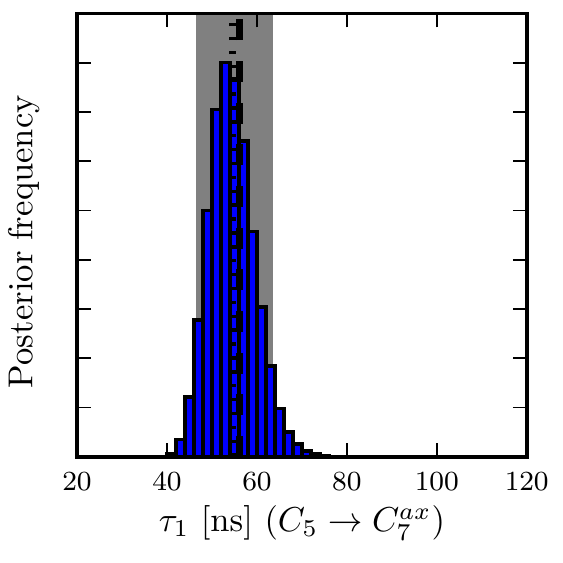}
(e) \includegraphics[width=0.25\textwidth]{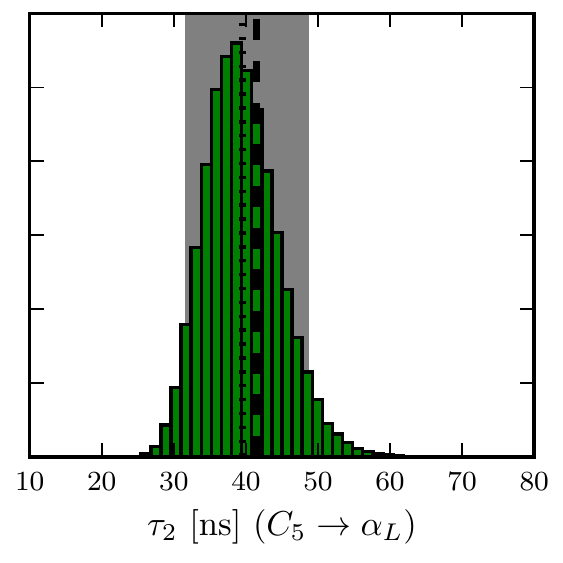}
(f) \includegraphics[width=0.25\textwidth]{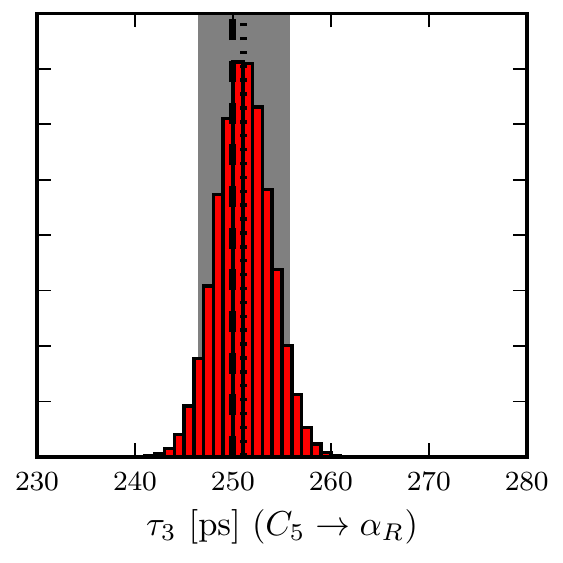}

\protect\caption{\label{fig:tsrevpi}a)-c) Implied time-scales, $t_{i}$. Histograms
obtained from reversible posterior sampling with fixed stationary
vector. Dashed lines indicate the  reference value, $\hat{t}_{i}$,
dotted lines indicate the posterior sample mean, $\mu(t_{i})$. The
90\% credible intervals are the shaded regions in gray.  d)-f) Expected
hitting times $\tau$. Histograms obtained from reversible posterior
sampling with fixed stationary vector. Dashed lines indicate the 
reference value, $\hat{\tau}$, dotted lines indicate the posterior
sample mean, $\mu(\tau)$.  The reference value obtained from a long
simulation is clearly compatible with the posterior sample (credible
interval) in all cases except the .}
\end{figure*}

Results are shown in In Table \ref{tab:ts_rev_pi} and Fig. \ref{fig:tsrevpi}.
The sample mean is again in good agreement with the  reference value.
For the computatio of the reference value we use the MLE transition
matrix of the full simulation data reversible with respect to the
input stationary distribution for the posterior sampling. The additional
constraint imposed by fixing the stationary distribution is clearly
reflected in smaller standard deviations for all shown observables
compared to the reversible case.

\begin{table}
\begin{ruledtabular}
\begin{tabular}{rrrrrr}
 & $\hat{t}_{i}/ps$  & $\mu/ps$  & $\sigma/ps$  & $\epsilon/ps$  & $t_{corr}$ \tabularnewline
$t_{2}$  &   1594 & 1520  & 196  & 0.6  & 1 \tabularnewline
$t_{3}$  & 72 & 73  & 1  & 0.003  & 1 \tabularnewline
$t_{4}$  &  38 & 41  & 3  & 0.01  & 1 \tabularnewline
\end{tabular}
\end{ruledtabular}

\protect\caption{\label{tab:ts_rev_pi} Comparison of reference implied time-scales
, ($\hat{t}_{i}$), with mean $\mu$ and standard deviation $\sigma$
from the reversible posterior using a fixed equilibrium distribution
and $N=10^{5}$ samples. $\epsilon$ is the estimated error of the
mean $\mu$ and $t_{\mathrm{corr}}$ is the autocorrelation time of
the sampled quantity.}
\end{table}

Histograms for expected hitting times between meta-stable sets are
shown in Fig. \ref{fig:tsrevpi}d-f). The sample mean is again in
good agreement with the  reference value. Again, we summarize our
results, c.f. Table \ref{tab:mfpt_rev_pi}. 

\begin{table}
\begin{ruledtabular}
\begin{tabular}{rrrrrr}
 & $\hat{\tau}/ns$  & $\mu/ns$  & $\sigma/ns$  & $t_{corr}$  & $\epsilon/ns$ \tabularnewline
$\tau(C_{5}\rightarrow C_{7}^{ax})$  &  56.0 & 54.5  & 5.0  & 1  & 0.02 \tabularnewline
$\tau(C_{5}\rightarrow\alpha_{L})$  & 41.5 & 39.5  & 5.1  & 1  & 0.02 \tabularnewline
$\tau(C_{5}\rightarrow\alpha_{R})$  & 0.249  & 0.251  & 0.003  & 1  & $9.7\cdot10^{-6}$ \tabularnewline
\end{tabular}
\end{ruledtabular}

\protect\caption{\label{tab:mfpt_rev_pi} Expected hitting times computed from reversible
sampling with fixed stationary vector using $N=10^{5}$ samples. See
Table \ref{tab:ts_rev_pi} for the definition of symbols.}
\end{table}

\subsubsection{Bovine pancreatic trypsin inhibitor, reversible sampling}

For BPTI, we used the 1 ms simulation generated on the Anton supercomputer
\cite{Shaw_Science10_Anton}. Please refer to that paper for system
setup and simulation details. We prepared data as follows: $C_{\alpha}$
atom positions were oriented to the mean structure and saved every
10 ns, resulting in about 100,000 configurations with 174 dimensions.
Time-lagged independent component analysis (TICA) \cite{PerezEtAl_JCP13_TICA,SchwantesPande_JCTC13_TICA}
was applied to reduce this 174-dimensional space to the two dominant
IC's as a spectral gap was found after the second nontrivial eigenvalue.
$k$-means clustering with $k=100$ was used to discretize this space. 

\begin{figure}
\includegraphics[width=1\columnwidth]{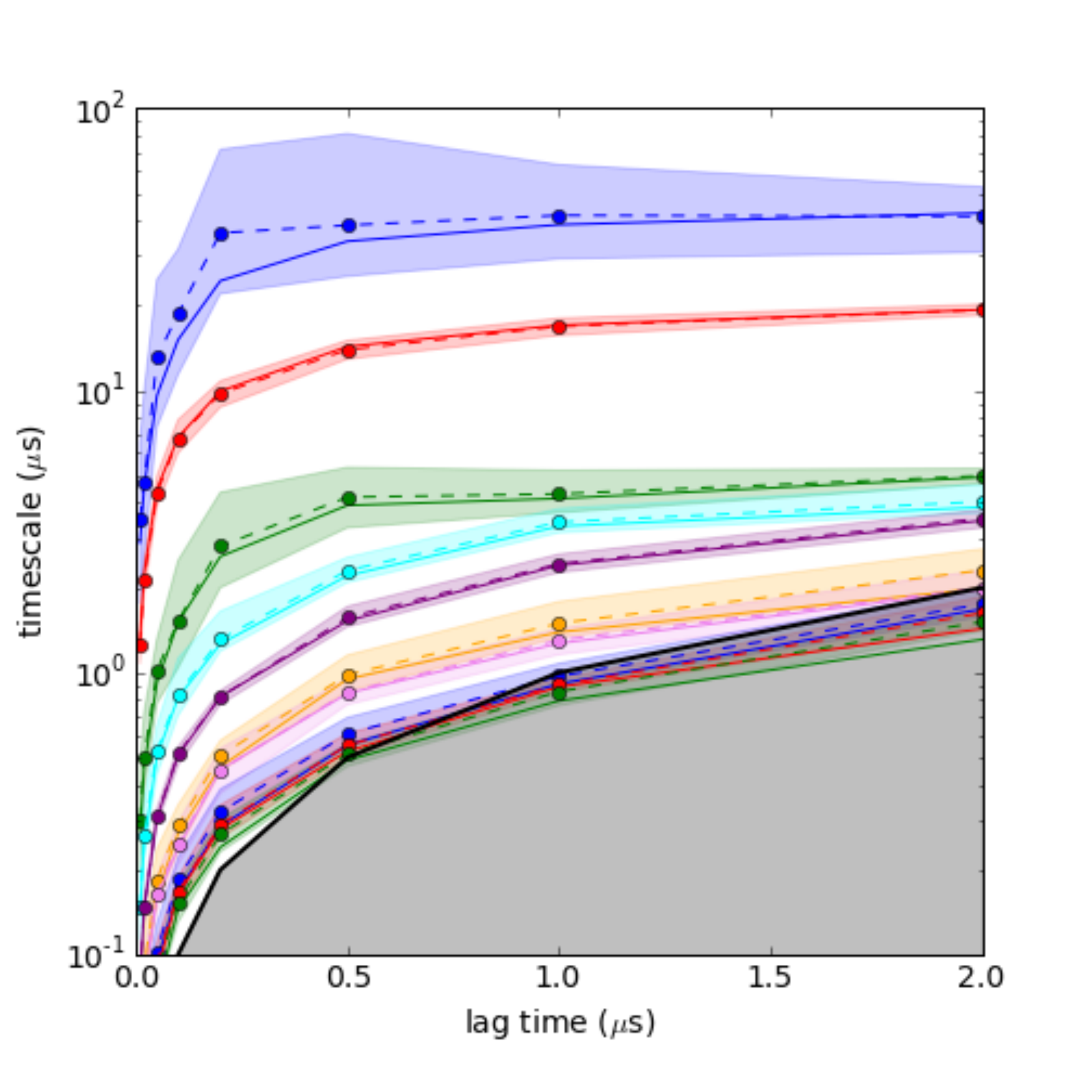}

\protect\caption{\label{fig:bpti_its}Implied timescales for a Markov model of bovine
pancreatic trypsin inhibitor (BPTI). The error bars are 95\% confidence
intervals estimated using the reversible transition matrix sampling
algorithm described here using transition counts as described in Ref.
\cite{Noe_preprint15_EffectiveCountMatrix}.}
\end{figure}

Effective count matrices were obtained using the method described
in \cite{Noe_preprint15_EffectiveCountMatrix} at a range of lag times
up to 2 $\mu$s. Fig. \ref{fig:bpti_its} shows the implied relaxation
timescales obtained from a maximum likelihood estimate with values
comparable to the Hidden Markov model analysis in Ref. \cite{NoeEtAl_PMMHMM_JCP13}.
The figure also shows uncertainties computed from a reversible transition
matrix sampling as described above with $N=1000$ samples. Only every
$20$th transition matrix sample was used to compute timescales in
order to reduce the computational effort to $50$ eigenvalue decompositions.
It is seen that the MLE is nicely contained in the 2$\sigma$ (95\%)
credible interval. The entire transition matrix sampling for Fig.
\ref{fig:bpti_its} took about 12.5 seconds on a 1.7 GHz Intel Core
i7. Given that 8 lag times were used for $1000$ samples of $100\times100$
matrices that contained about 40\% non-zeros, about 2.56 million elements
are sampled per second, and about 640 full transition matrix samples
are generate per second. Below a more systematic analysis of the computational
efficiency is made.

\FloatBarrier

\subsection{Efficiency}

We compute acceptance probabilities of the Metropolis-Hastings steps
and compare the statistical efficiency of the proposed sampling algorithm
with the algorithm proposed in Ref. \cite{noe2008} that uses uniform
proposal densities. Efficiency is measured in terms of achieved autocorrelation
times for sampling of transition matrices with different sizes. As
a representative observable we choose the largest relaxation time-scale
$t_{2}$ for the alanine dipeptide molecule and compute autocorrelation
functions and autocorrelation times from a large sample of size $N=10^{6}$.
Two differently fine discretizations were used, resulting in $n\times n$-shaped
transition matrices with $n=258$ and $n=1108$.

\begin{figure}
(a) \includegraphics[width=0.8\columnwidth]{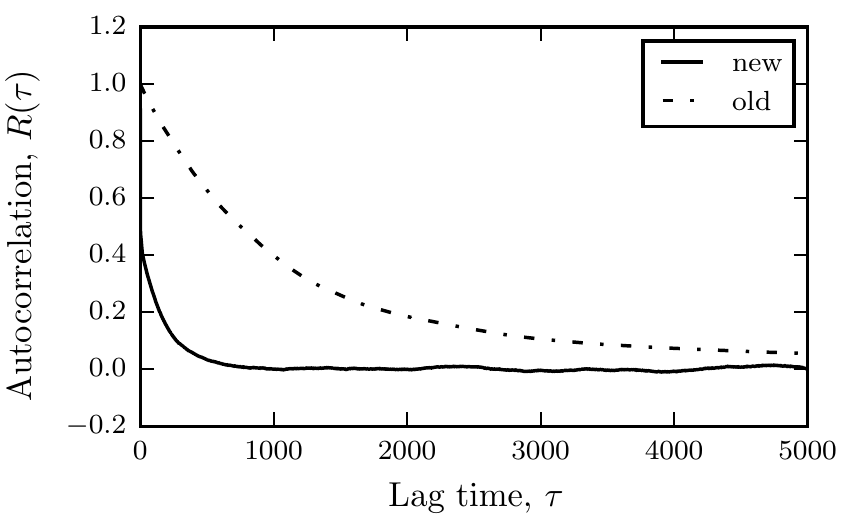}

(b) \includegraphics[width=0.8\columnwidth]{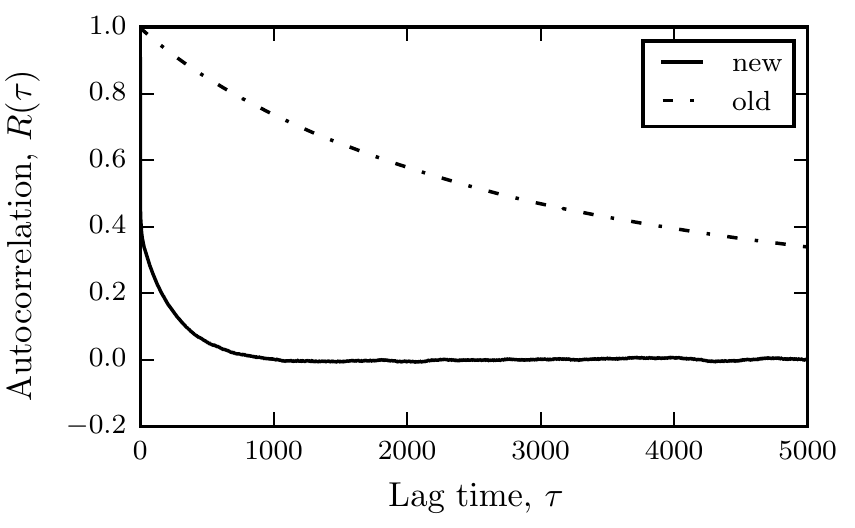}

\protect\caption{\label{fig:acf_ts_rev_old_vs_new}Autocorrelation function for reversible
sampling. Sampling of transition matrices with $n=233$ states, a),
and b) with $n=1108$ states. Increasing the dimension of the state
space has only a small effect on the new sampling algorithm.}
 
\end{figure}

Fig. \ref{fig:acf_ts_rev_old_vs_new} shows autocorrelation functions
for the second largest relaxation time-scale, $t_{2}$, for a reversible
posterior ensemble. The autocorrelation function for the reversible
sampling algorithm with posterior adapted proposals shows a much faster
decay than the autocorrelation function for the algorithm in Ref.
\cite{noe2008}. Table \ref{tab:efficiency_rev} compares acceptance
probabilities and autocorrelation times. The present proposal steps
lead to very high acceptance probabilities, $p>0.99$, for the sampling
off-diagonal entries. The main advance, however comes from the fact
that the step for sampling diagonal transition matrix elements in
Ref. \cite{noe2008} has suffered from a very poor acceptance probability.
As that step was the only step that modified the equilibrium distribution,
the sampler in Ref. \cite{noe2008} has very poor mixing properties.
In contrast, our current algorithm generates independent samples for
the diagonal elements, resulting in an acceptance probability of $p=1.0$. 

The autocorrelation times for the new sampler are more than a factor
5 shorter for the small (233 state) matrix and more than a factor
13 shorter for the large (1108 state) matrix, indicating a much higher
efficiency of the new approach. The autocorrelation time of the new
algorithm increases only mildly for matrices of increased dimension,
indicating that the present algorithm will be useful for very large
Markov models. 

\begin{table}
\begin{ruledtabular}
\begin{tabular}{rrrrr}
algorithm  & $n$  & $p_{\text{offdiag}}$  & $p_{\text{diag}}$  & $t_{\text{corr}}$ \tabularnewline
\multirow{2}{*}{old} & 233  & 0.216  & 0.011  & 1088.1 \tabularnewline
 & 1108  & 0.271  & 0.005  & 3241.9 \tabularnewline
\multirow{2}{*}{new} & 233  & 0.994  & 1.0  & 194.7 \tabularnewline
 & 1108  & 0.995  & 1.0  & 242.6 \tabularnewline
\end{tabular}
\end{ruledtabular}

\protect\caption{\label{tab:efficiency_rev}Acceptance probability and autocorrelation
times for old vs. new reversible sampling algorithm. $n$: number
of states; $p_{\mathrm{offdiag}}$, $p_{\mathrm{diag}}$: acceptance
probabilities for off-diagonal and diagonal elements, respectively.
$t_{\mathrm{corr}}$: Autocorrelation time for the sampling of the
slowest relaxation timescale, in number of transition matrix sampling
steps.}
 
\end{table}

\begin{figure}
(a) \includegraphics[width=0.8\columnwidth]{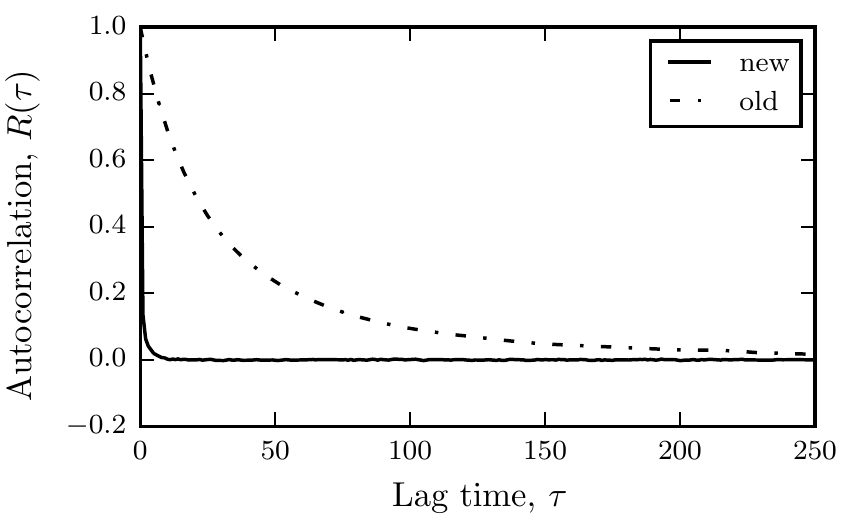}

(b) \includegraphics[width=0.8\columnwidth]{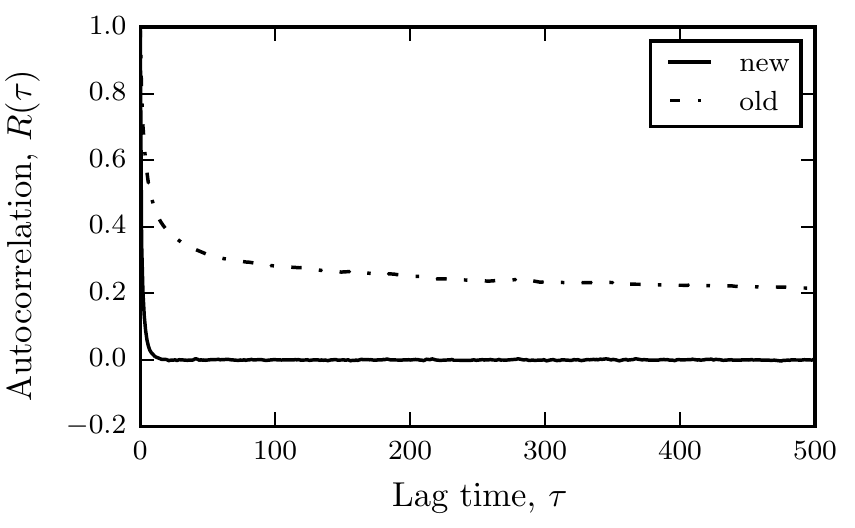}

\protect\caption{\label{fig:acf_ts_revpi_vs_new}Autocorrelation function for reversible
sampling with fixed stationary distribution. Sampling of transition
matrices with $n=233$ states, a), and b) with $n=1108$ states. Increasing
the dimension of the state space has only a small effect on the new
sampling algorithm.}
 
\end{figure}

Fig. \ref{fig:acf_ts_revpi_vs_new} shows autocorrelation functions
for reversible sampling with fixed stationary vector. The posterior
adapted proposals in reversible sampling algorithm with fixed stationary
distribution again result in a much faster decay of the autocorrelation
function than the uniform proposals of the algorithm in Ref. \cite{noe2008}.
Table \ref{tab:efficiency_revpi} compares acceptance probabilities
and autocorrelation times for the two algorithms. For sampling with
fixed stationary vector there is no sampling step for the diagonal
elements. Although the average acceptance probabilities are only a
factor of 3-4 better for our new algorithm, the autocorrelation times
are decreased by a factor 35 for the small system (233 states) and
over a factor 300 for the large system (1108 states). Again, there
is only a  mild increase in autocorrelation time when the dimension
of the sampled space is increased. 

\begin{table}
\begin{ruledtabular}
\begin{tabular}{rrrr}
algorithm  & $n$  & $p_{\text{offdiag}}$  & $t_{\text{corr}}$ \tabularnewline
\multirow{2}{*}{old} & 233  & 0.175  & 72.096 \tabularnewline
 & 1108  & 0.230  & $>1000$ \footnote{Autocorrelation function not converged} \tabularnewline
\multirow{2}{*}{new} & 233  & 0.752  & 2.893 \tabularnewline
 & 1108  & 0.706  & 3.157 \tabularnewline
\end{tabular}
\end{ruledtabular}

\protect\caption{\label{tab:efficiency_revpi}Acceptance probability and autocorrelation
times for old vs. new reversible sampling algorithm with fixed stationary
distribution. Symbols as in Table \ref{tab:efficiency_rev}.}
 
\end{table}

\section{Conclusion}

In this work we have described and significantly extended the state
of the art in reversible Markov model estimation. Reversible Markov
models are expected to naturally arise from molecular dynamics implementations
that fulfill microscopic reversibility. Reversibility is an essential
property in order to analyze the equilibrium kinetics of a molecule.
However, in order to have reversibility in a Markov model, it needs
to be enforced in the estimation procedure. When done correctly, reversible
estimation does not bias the model but rather reduces statistical
errors as a result of a smaller number of degrees of freedom.

We have presented minor improvements to an existing self-consistent
estimation algorithm for reversible Markov models. Then we have presented
a new and efficient algorithm to estimate reversible maximum likelihood
Markov models given a fixed equilibrium distribution.

The main part of the presented work focuses on the long-standing problem
of Bayesian estimation of the posterior ensemble of reversible transition
matrices. Although several algorithms to sample reversible Markov
models have been presented in the past, they have been hampered by
three fundamental problems, two of which are addressed here: (i) Which
prior should be chosen such that the posterior is located around the
 true value rather than completely elsewhere? (ii) How should transition
counts be obtained when time series are correlated and not really
Markovian at any given lag time $\tau$? (iii) How can the sampling
algorithm be made efficient such that also large transition matrices
can be sampled in reasonable time?

To address problem (i), we develop priors that ensure that  reference
values and sample mean are similar. The key property of these priors
is that they make the \emph{a priori} assumption that transitions
between states that have not been sampled in the trajectory in either
direction, have zero probability. This is a sparse prior, i.e. an
improper prior enforcing that the sampled transition matrix has the
same sparsity structure as the maximum likelihood estimate and as
induced by the observation. In contrast to most other priors that
have been previously suggested, these sparse priors achieve the desired
property of creating errors bars that nicely envelop the reference
estimates.

For problem (ii), we have described the principles of how it can be
addressed. We suggest that effective count matrices are obtained using
the concept of statistical inefficients. A separate preprint \cite{Noe_preprint15_EffectiveCountMatrix}
suggests an initial solution towards this aim that is successfully
applied on simulation data of the bovine pancreatic trypsin inhibitor
in the present paper. The solution of problem (ii) is still in its
infancy and needs further investigation..

For problem (iii), we present highly efficient Gibbs sampling algorithms
for reversible transition matrices and reversible transition matrices
with fixed equilibrium distribution. Both methods are demonstrated
to have acceptance probabilities close to 1 in their individual update
steps. Autocorrelation times from samples of the slowest relaxation
timescale are one or two orders of magnitude shorter than with a previous
Gibbs sampling algorithm, indicating a high statistical efficiency
of our sampler.

Implementations of all algorithms described here are available in
PyEMMA \cite{SchererEtAl_JCTC15_EMMA2} as of version 2.0 or later
 - \href{http://www.pyemma.org}{www.pyemma.org} .


\begin{acknowledgements} This work was funded by the Deutsche Forschungsgemeinschaft
(DFG) projects SFB 1114/C03 (FP), SFB 1114/A04 (HW), grant NO 825/3-1
(BTS) and the European Research Council (ERC) starting grant ``pcCell''
(FN, BTS, FB). \end{acknowledgements}


%

\clearpage


\appendix

\section{Details for transition matrix sampling}

\subsection{Reversible transition matrix sampling: derivation of marginal densities}

We first pick a single element $(k,\,l)$ of $X$ (diagonal or off-diagonal)
and sample it from a proposal density $x_{kl}^{\prime}\sim q(x_{kl}^{\prime}|X)$
that is scale-invariant with $q(x_{kl}^{\prime}|X)\propto q(cx_{kl}^{\prime}|cX)$
for all $c>0$:
\begin{equation}
x_{ij}^{\prime}=\left\{ \begin{array}{ll}
\sim q(x_{kl}^{\prime}|X), & (i,j)=(k,l)\\
x_{ij}, & \text{else}
\end{array}\right.
\end{equation}
and then renormalize the matrix such that it retains an element sum
of 1:
\begin{equation}
\bar{x}_{ij}^{\prime}=\frac{x_{ij}}{1-x_{kl}+x_{kl}^{\prime}}\label{eqapx:x-bar-prime}
\end{equation}
Since $q(x_{kl}^{\prime}|X)$ is a probability density function, we
can obtain from its scale-invariance that
\begin{eqnarray}
\int q(cx_{kl}^{\prime}|cX)\mathrm{d}x_{kl}^{\prime} & = & \frac{1}{c}\int q(cx_{kl}^{\prime}|cX)\mathrm{d}\left(cx_{kl}^{\prime}\right)=\frac{1}{c}\nonumber \\
\end{eqnarray}
and
\begin{equation}
q(cx_{kl}^{\prime}|cX)=\frac{1}{c}q(x_{kl}^{\prime}|X)
\end{equation}
According to Theorem 13.1 in \cite{Sawyer_Book06_MetropolisHastings},
the posterior distribution $\mathbb{P}(X|C)$ is the invariant distribution
of the proposed update step if we accept $\bar{X}^{\prime}$ as the
new sample with probability $\min\{1,p_{acc}\}$ and
\begin{equation}
p_{acc}=\frac{\mathbb{P}(\bar{X}^{\prime})}{\mathbb{P}(X)}\cdot\frac{\mathbb{P}(C|\bar{X}^{\prime})}{\mathbb{P}(C|X)}\cdot\frac{q_{x}(x_{ij}|\bar{X}^{\prime})}{q_{x}(\bar{x}_{ij}^{\prime}|X)}\cdot\prod_{(i,j)\neq(k,l),(i^{\prime},j^{\prime})}\frac{\partial\bar{x}_{ij}^{\prime}}{\partial x_{ij}}\label{eq:p_acc}
\end{equation}
where $q_{x}(\bar{x}_{kl}^{\prime}|X)$ denotes the proposal density
of $\bar{x}_{ij}^{\prime}$ given $X$. Note $X$ only contains $n(n+1)/2-1$
free variables. So we select $\{x_{ij}|i\ge j,(i,j)\neq(i^{\prime},j^{\prime})\}$
as the free variable set of $X$, where $x_{i^{\prime}j^{\prime}}$
is an arbitrary element of $X$ with $i^{\prime}\ge j^{\prime}$ and
$(i^{\prime},j^{\prime})\neq(k,l)$. Let us consider each term on
the right hand side of \eqref{eq:p_acc}.

From the definition of $\bar{X}^{\prime}$, we have

\begin{eqnarray}
\frac{\mathbb{P}(\bar{X}^{\prime})}{\mathbb{P}(X)} & = & \left(\frac{1}{1-x_{kl}+x_{kl}^{\prime}}\right)^{b_{0}}\left(\frac{y}{x_{kl}}\right)^{b_{kl}}\nonumber \\
 & = & \left(\frac{1-\bar{x}_{kl}^{\prime}}{1-x_{kl}}\right)^{b_{0}}\left(\frac{x_{kl}^{\prime}}{x_{kl}}\right)^{b_{kl}}
\end{eqnarray}
and
\begin{equation}
\frac{\mathbb{P}(C|\bar{X}^{\prime})}{\mathbb{P}(C|X)}=\left\{ \begin{array}{ll}
\frac{\left(x_{kk}^{\prime}\right)^{c_{kk}}(x_{k}-x_{kk}+x_{kk}^{\prime})^{-c_{k}}}{x_{kk}^{c_{kk}}(x_{k}-x_{kk}+x_{kk})^{-c_{k}}}, & k=l\\
\frac{\left(x_{kl}^{\prime}\right)^{c_{kl}+c_{lk}}(x_{k}-x_{kl}+x_{kl}^{\prime})^{-c_{k}}}{x_{kl}^{c_{kl}+c_{lk}}(x_{k}-x_{kl}+x_{kl})^{-c_{k}}}\\
\times\frac{(x_{l}-x_{kl}+x_{kl}^{\prime})^{-c_{l}}}{(x_{l}-x_{kl}+x_{kl})^{-c_{l}}}, & k\neq l
\end{array}\right.
\end{equation}
The proposal density of $\bar{x}_{kl}^{\prime}$ given $X$ can be
expressed as
\begin{eqnarray}
q_{x}(\bar{x}_{kl}^{\prime}|X) & = & \left(\frac{\partial\bar{x}_{kl}^{\prime}}{\partial x_{kl}^{\prime}}\right)^{-1}q(x_{kl}^{\prime}|X)\nonumber \\
 & = & \frac{1-x_{kl}}{(1-\bar{x}_{kl}^{\prime})^{2}}q(x_{kl}^{\prime}|X)\nonumber \\
 & = & \frac{1-x_{kl}}{(1-\bar{x}_{kl}^{\prime})^{2}}q\left(\frac{1-x_{kl}}{1-\bar{x}_{kl}^{\prime}}\cdot\bar{x}_{kl}^{\prime}|X\right)
\end{eqnarray}
Therefore,
\begin{eqnarray}
q_{x}(x_{ij}|\bar{X}^{\prime}) & = & \frac{1-\bar{x}_{kl}^{\prime}}{(1-x_{kl})^{2}}q\left(\frac{1-\bar{x}_{kl}^{\prime}}{1-x_{kl}}\cdot x_{kl}|\bar{X}^{\prime}\right)\nonumber \\
 & = & \frac{1}{1-x_{kl}}q\left(x_{kl}|X^{\prime}\right)
\end{eqnarray}
and
\begin{equation}
\frac{q_{x}(x_{ij}|\bar{X}^{\prime})}{q_{x}(\bar{x}_{kl}^{\prime}|X)}=\left(\frac{1-x_{kl}}{1-\bar{x}_{kl}^{\prime}}\right)^{2}\frac{q(x_{kl}^{\prime}|X)}{q(x_{kl}|X^{\prime})}
\end{equation}
The partitial derivatitive of $\bar{x}_{ij}^{\prime}$ with respect
to $x_{ij}$ for $(i,j)\neq(k,l)$ can be calculated according to
\eqref{eqapx:x-bar-prime} as
\begin{equation}
\frac{\partial\bar{x}_{ij}^{\prime}}{\partial x_{ij}}=\frac{1}{1-x_{kl}+x_{kl}^{\prime}}=\frac{1-\bar{x}_{kl}^{\prime}}{1-x_{kl}}
\end{equation}
Combining all the above results leads to
\begin{equation}
p_{acc}=\left(1-x_{kl}+x_{kl}^{\prime}\right)^{-\frac{n(n+1)}{2}-b_{0}}\frac{q(x_{kl}|X^{\prime})}{q(x_{kl}^{\prime}|X)}\frac{\gamma(x_{kl}^{\prime}|X)}{\gamma(x_{kl}|X)}
\end{equation}
with
\begin{equation}
\gamma(x_{kl}^{\prime}|X)\propto\left\{ \begin{array}{ll}
\frac{\left(x_{kk}^{\prime}\right)^{c_{kk}+b_{kk}}}{(x_{k}-x_{kk}+x_{kk}^{\prime})^{c_{k}},} & k=l\\
\frac{\left(x_{kl}^{\prime}\right)^{c_{kl}+c_{lk}+b_{kl}}}{(x_{k}-x_{kl}+x_{kl}^{\prime})^{c_{k}}(x_{l}-x_{kl}+x_{kl}^{\prime})^{c_{l}},} & k\neq l
\end{array}\right.
\end{equation}

\subsection{Reversible transition matrix sampling: Efficient proposal densities}

\paragraph{Diagonals: }

Let us define variable $s^{\prime}=\frac{x_{kk}'}{x_{k}-x_{kk}+x_{kk}'}$.
If $x_{kk}^{\prime}$ is sampled from the proposal density $\gamma(x_{kk}^{\prime}|X)$,
the corresponding proposal density of $s^{\prime}$ can be expressed
as
\begin{equation}
s^{\prime}\sim\left|\frac{\partial x_{kk}^{\prime}}{\partial s^{\prime}}\right|\gamma(x_{kk}^{\prime}|X)
\end{equation}
Note that
\begin{equation}
x_{kk}^{\prime}=(x_{k}-x_{kk})\frac{s^{\prime}}{1-s^{\prime}}\label{eq:back-transform-xkk-s}
\end{equation}
Therefore
\begin{equation}
\frac{\partial x_{kk}^{\prime}}{\partial s^{\prime}}=\left(x_{k}-x_{kk}\right)\left(1-s^{\prime}\right)^{-2}
\end{equation}
and
\begin{eqnarray}
s^{\prime} & \sim & (x_{k}-x_{kk})\left(1-s^{\prime}\right)^{-2}\cdot\left(x_{kk}^{\prime}\right)^{c_{kk}-1}\cdot(x_{k}-x_{kk}+x_{kk}^{\prime})^{-c_{k}}\nonumber \\
 & \propto & \left(1-s^{\prime}\right)^{-2}\cdot\left(\frac{s^{\prime}}{1-s^{\prime}}\right)^{c_{kk}-1}\cdot\left(1-s^{\prime}\right)^{c_{k}}\nonumber \\
 & = & \left(s^{\prime}\right)^{c_{kk}-1}\left(1-s^{\prime}\right)^{c_{k}-c_{kk}-1}
\end{eqnarray}
The above equation implies that $s^{\prime}$ follows the Beta distribution
with parameters $c_{kk}$ and $c_{k}-c_{kk}$. So we can sample $s^{\prime}\sim\mathrm{Beta}(c_{kk},c_{k}-c_{kk})$
and get $x_{kk}^{\prime}$ by the back-transform.

\paragraph{Off-diagonals: }

We consider how to approximate $\gamma(x_{kl}^{\prime}|X)$ with $k\neq l$
and $b_{kl}=-1$ by a gamma distribution density function. Define
\begin{align}
\left(x_{kl}^{\prime}\right)^{c_{kl}+c_{lk}-1}(x_{k}-x_{kl}+x_{kl}^{\prime})^{-c_{k}}(x_{l}-x_{kl}+x_{kl}^{\prime})^{-c_{l}}\\
=\left(x_{kl}^{\prime}\right)^{-1}\exp f(x_{kl}^{\prime}) & .
\end{align}
The function $f(x_{kl}^{\prime})$ is then given by 
\begin{equation}
\begin{aligned}f(x_{kl}^{\prime})=(c_{kl}+c_{lk})\log x_{kl}^{\prime}-c_{k}\log(x_{k}-x_{kl}+x_{kl}^{\prime})\\
-c_{l}\log(x_{l}-x_{kl}+x_{kl}^{\prime}).
\end{aligned}
\label{eq:conditional_vkl_f}
\end{equation}
We approximate $f$ using a three parameter family of functions 
\begin{equation}
\hat{f}(x_{kl}^{\prime}|\alpha,\beta,f_{0})=\alpha\log x_{kl}^{\prime}-\beta x_{kl}^{\prime}+f_{0}\label{eq:approximation_f}
\end{equation}
so that the corresponding approximate $\gamma(x_{kl}^{\prime}|X)$
is 
\begin{equation}
\hat{\gamma}(x_{kl}^{\prime}|X)\propto\left(x_{kl}^{\prime}\right)^{-1}\exp\hat{f}(x_{kl}^{\prime}|\alpha,\beta,f_{0}).\label{eq:approximation_conditional}
\end{equation}
\eqref{eq:approximation_conditional} is a Gamma distribution with
parameters $\alpha$, $\beta$ which can be efficiently sampled \cite{devroye1986}.

The three parameters $\alpha$, $\beta$, $f_{0}$ are obtained matching
$f$ and $\hat{f}$ up to second derivatives at the maximum point
\begin{equation}
\bar{v}=\arg\max f(x_{kl}^{\prime}).\label{eq:maximum_point_def}
\end{equation}
This leads to the following linear system 
\begin{subequations}
\begin{align}
\log\bar{v}\,\alpha+\bar{v}\beta+f_{0} & =f(\bar{v})\\
\frac{\alpha}{\bar{v}}-\beta & =f'(\bar{v})=0\\
-\frac{\alpha}{\bar{v}^{2}} & =f''(\bar{v})=h
\end{align}
with solution 

\begin{align}
\alpha & =-h\bar{v}^{2}\label{eqapx:alpha}\\
\beta & =-h\bar{v}\label{eqapx:beta}\\
f_{0} & =f(\bar{v})+h\bar{v}^{2}(\log\bar{v}-1)\label{eqapx:f0}
\end{align}
and
\[
h=f''(\bar{v})=\frac{c_{k}}{(x_{k}-x_{kl}+\bar{v})^{2}}+\frac{c_{l}}{(x_{l}-x_{kl}+\bar{v})^{2}}-\frac{c_{kl}+c_{lk}}{\bar{v}^{2}}
\]
The maximum point can be computed as the root of a quadratic equation
in the usual way, 
\begin{equation}
\bar{v}=\frac{-b+\sqrt{b^{2}-4ac}}{2a}\label{eq:maximum_point}
\end{equation}
with parameters $a$, $b$, $c$ given by 

\begin{align}
a & =c_{k}+c_{l}-c_{kl}-c_{lk}\label{eqapx:a}\\
b & =(c_{k}-c_{kl}-c_{lk})(x_{l}-x_{kl})\label{eqapx:b}\\
 & \:\:\:+(c_{l}-c_{kl}-c_{lk})(x_{k}-x_{kl})\\
c & =-(c_{kl}+c_{lk})(x_{k}-x_{kl})(x_{l}-x_{kl})\label{eqapx:c}
\end{align}

\end{subequations}

The second solution corresponding to \eqref{eq:maximum_point} with
negative sign in front of the square root can be safely excluded since
$\bar{y}$ is required to be non-negative.

\subsection{Reversible sampling with fixed stationary distribution: efficient
proposal densities}

We first prove that there exists a maximum likelihood estimate $\hat{\mathbf{P}}=[\hat{p}_{ij}]$
of the transition matrix which satisfies $|\{k|\hat{p}_{kk}>0,c_{kk}=0\}|\le1$.
Suppose that $\mathbf{X}^{\prime}=[x_{ij}^{\prime}]$ is a maximum
likelihood estimate of $\mathbf{X}$ and there are two different indices
$k,l$ which satisfy that $x_{ll}^{\prime}\ge x_{kk}^{\prime}>0$
and $c_{kk}=c_{ll}=0$. We can then construct a new matrix $\mathbf{X}^{\prime\prime}=[x_{ij}^{\prime\prime}]$
with
\begin{align*}
x_{kl}^{\prime\prime} & =x_{kl}^{\prime}+x_{kk}^{\prime}\\
x_{kk}^{\prime\prime} & =0\\
x_{kl}^{\prime\prime} & =x_{kl}^{\prime}+x_{kk}^{\prime}\\
x_{ll}^{\prime\prime} & =x_{ll}^{\prime}+x_{kl}^{\prime}-x_{kk}^{\prime}.
\end{align*}
It can be verified that the likelihood of $\mathbf{X}^{\prime\prime}$
is larger than $\mathbf{X}^{\prime}$, which implies that $\mathbf{X}^{\prime\prime}$
is also a maximum lielihood estimate of $\mathbf{X}$. Repeat the
above procedure, we can finally get an maximum likelihood estimate
$\hat{\mathbf{X}}$ of $\mathbf{X}$ which has at most one $k$ with
$\hat{x}_{kk}>0$ and $c_{kk}=0$, and the corresponding $\hat{\mathbf{P}}$
satisfies that $|\{k|\hat{p}_{kk}>0,c_{kk}=0\}|\le1$.

We will now investigate how to approximate the density 
\begin{equation}
\gamma_{v}(v^{\prime}|X)\propto\left(v^{\prime}\right)^{a_{1}}\left(\frac{s}{s-1}+v^{\prime}\right)^{a_{3}}\bigg(1+v^{\prime}\bigg)^{-(a_{1}+a_{2}+a_{3}+2)}.\label{eq:density_v_fixed_pi}
\end{equation}
with $s>1$ by a Gamma distribution density.

As in the reversible case we will use the representation 
\begin{equation}
\gamma_{v}(v^{\prime}|X)=\left(v^{\prime}\right)^{-1}\exp f(v^{\prime})\label{eq:eq:density_v_fixed_pi_exp}
\end{equation}
and approximate $f(v^{\prime})$ using the three parameter family,
$\hat{f}(v^{\prime}|\alpha,\beta,f_{0})$, given in \eqref{eq:approximation_f}.
The resulting approximate density, $\hat{\gamma}_{v}(v^{\prime}|X)$,
has the same nice properties as the one from \eqref{eq:approximation_conditional}.

Parameters $\alpha$ $\beta$ and $f_{0}$ are given by \eqref{eqapx:alpha},
\eqref{eqapx:beta} and \eqref{eqapx:f0}. The maximum point $\bar{v}$
is given by \eqref{eq:maximum_point}. The parameters $a$, $b$,
$c$ are, 
\begin{subequations}
\begin{align}
a & =a_{2}+1\label{eq:a_fixed_pi}\\
b & =a_{2}-a_{1}+\frac{a_{2}+a_{3}+1}{s-1}\label{eq:b_fixed_pi}\\
c & =\frac{s(a_{1}+1)}{1-s}\label{eq:c_fixed_pi}
\end{align}
\end{subequations}

\end{document}